\documentclass[12pt]{article}

\usepackage{amsmath,amssymb,amsthm}
\usepackage[margin=2.5cm]{geometry}
\usepackage{hyperref}
\usepackage{color}
\usepackage{graphicx}
\usepackage{pgf}

\newcounter{Theorems}
\newcounter{Definitions}
\newcounter{Conjectures}
\usepackage{array}
\usepackage{diagbox}

\begin{document}

\begin{titlepage}
  \begin{flushright}

  \end{flushright}

  \begin{center}
    {\Large\bf $ $ \\ $ $ \\
    Supersymmetry bicomplex of pure spinor AdS background
    } \\
    \bigskip\bigskip\bigskip
    {\large Thiago Oliveira Ferreira}\\
    \bigskip
        {\it Perimeter Institute for Theoretical Physics\\
          31 Caroline St. N.,\\
          Waterloo, Ontario, Canada, N2L 2Y5
            }
            \bigskip
            and
            \\
            \bigskip
                {\large Andrei Mikhailov}
                \\
                \bigskip
                    {\it Instituto de F\'{i}sica Te\'{o}rica, Universidade Estadual Paulista\\
                      R. Dr. Bento Teobaldo Ferraz 271, 
                      Bloco II -- Barra Funda\\
                      CEP: 01140-070 -- S\~{a}o Paulo, Brasil\\
                    }

                    \vskip 1cm
  \end{center}

\begin{abstract}
Infinitesimal deformations of $\text{AdS}_5 \times \mathbb{S}^5$ form a representation of the AdS supersymmetry algebra.
The structure of this representation has not yet been completely described in the literature.
Some information can be obtained just from the fact that the space of deformations is the cohomology of a nilpotent BRST operator. We can consider the bicomplex formed by the BRST operator and the Lie cohomology differential, and its two spectral sequences.
Their matching imposes some constraints on the structure of representations, which we start exploring in this paper. In particular, we clarify the structure of ghost number
three zero modes.
\end{abstract}

  \vfill
      {\renewcommand{\arraystretch}{0.8}%
      }

\end{titlepage}

\tableofcontents

\section{Introduction}\label{Introduction}

A Type II supergravity background can be encoded as a nilpotent vector field $Q$ on the pure spinor cone bundle over space-time. Thus, to each background, there is an associated cohomological complex of $Q$, which is related to deformations of that background. Basically, in a generic background, the cohomology spaces in ghost numbers two and three correspond to infinitesimal deformations.

Special backgrounds with large supersymmetries, such as flat space or $\text{AdS}_5 \times \mathbb{S}^5$, are particularly interesting; in those backgrounds, one can study the action of symmetries on linearized deformations.
The space of linearized deformations is a representation of the supersymmetry algebra. If one admits non-normalizable deformations, then this representation is non-semisimple, i.e. it is \emph{not} a direct sum of irreducible representations. It is interesting to understand the structure of this representation. In $\text{AdS}_3$ this was done in \cite{Troost:2011fd, Gaberdiel:2011vf};
as far as we know, however, the complete description for $\text{AdS}_5$ is still missing.

In particular, one can compute the cohomology of the supersymmetry algebra with coefficients in the deformation representation. This corresponds to the ``Lie superalgebra cohomology complex'' (also known as Hochschild--Serre complex). This is yet another complex associated to a symmetric background, besides the complex of $Q$ aforementioned. Those two complexes can be combined into a bicomplex, and the two spectral sequences converging to the cohomology of the bicomplex. One spectral sequence, which we will denote by $E$, corresponds to first computing the cohomology of $Q$, yielding $H_Q \equiv H(Q)$, and then the Lie superalgebra cohomology with coefficients in $H_Q$. The other one, which will be denoted by $\widetilde{E}$, corresponds to computing first the Lie superalgebra cohomology, and then the cohomology of $Q$.

At the second page, the elements of $E_2$ are the cohomology groups of the supersymmetry algebra with values in the cohomology of $Q$. Moreover, there is a generalization of this construction for any representation $\mathcal{R}$, where
\begin{equation} \label{EvsExt}
    E^{p,q}_2 (\mathcal{R})
        = \operatorname{Ext}^p (\mathcal{R},H^q_Q) .
\end{equation}
This cohomology group is an important ``probe'' of the space of linearized solutions $H_Q$. Therefore, it is useful to know $E_2^{p,q}(\mathcal{R})$. Since both spectral sequences converge to the cohomology of the total complex, we have the matching condition
\begin{equation} \label{EisEtil}
    \bigoplus_{p+q=n} E_{\infty}^{p,q}
        = \bigoplus_{p+q = n} \widetilde{E}_{\infty}^{p,q} .
\end{equation}

Nonetheless, $\widetilde{E}$ is easier to compute than $E$. That is because the Lie superalgebra cohomology on off-shell vertices can be computed using Shapiro's lemma, which can be understood as a variation of the Poincar\'{e} lemma. Then $\widetilde{E}_2$ is the cohomology of $Q$ on the ``covariant subcomplex'', which is much smaller than the original pure spinor complex.

Note, however, that the relation between $E$ and $\widetilde{E}$ in equation \eqref{EisEtil} only works
at the limiting pages $E_\infty$ and $\widetilde{E}_\infty$ (also called ``$\infty$-th'' pages). Given equation \eqref{EvsExt}, we are interested in the second page $E_2$. As per usual in spectral sequences, the limiting page $E_{\infty}$ is obtained from $E_2$ by taking the cohomology with respect to $d_2$ and then higher differentials. Therefore, the $\operatorname{Ext}$ groups
in equation \eqref{EvsExt} cannot be obtained directly from $\widetilde{E}$. However, the matching condition \eqref{EisEtil} establishes interesting relations between those spectral sequences, which we will start unveiling in this paper.

This work is related to  \cite{Mikhailov:2009rx,Mikhailov:2019kfm,Mikhailov:2023fvx,Mikhailov:2024sef}.

\section{Supergeometry of \texorpdfstring{$\boldsymbol{\text{AdS}_5 \times  \mathbb{S}^5}$}{AdS5×S5}} \label{SupergeometryOfAdS}

\subsection{Notations}\label{sec:Notations}

We will start by introducing notations.

Suppose that $f$ is a linear function or a linear map between two linear spaces; we will then write
$f\langle x\rangle$ instead of $f(x)$ to remind ourselves that the dependence on $x$ is linear.

We sometimes use the notation $H_\mathfrak{g}(V)$ instead of $H(\mathfrak{g};V)$ (the cohomology of $\mathfrak{g}$ with coefficients in $V$), for brevity. $H_Q$ means cohomology of the pure spinor BRST operator.

\subsubsection{Supersymmetry algebra \texorpdfstring{$\mathfrak{g}$}{frak(g)}} \label{sec:SUSYAlgebra}

The superconformal algebra $\mathfrak{psu}(2,2|4)$ will be denoted simply by $\mathfrak g$, and the stabilizer of a point in $\text{AdS}_5 \times \mathbb{S}^5$ will be denoted $\mathfrak h$, viz.
\begin{align*}
    \mathfrak{g} &\equiv \mathfrak{psu}(2,2|4) , \\
    \mathfrak{h} &\equiv \mathfrak{so}(1,4) \oplus \mathfrak{so}(5) .
\end{align*}
One could also use larger Lie superalgebras, such as
\begin{alignat*}{3}
    & \mathfrak{g}' &&\equiv \mathfrak{pu}(2,2|4) ,
    \\
    & \widehat{\mathfrak{g}} &&\equiv \mathfrak{su}(2,2|4) ,
    \\
    & \widehat{\mathfrak{g}}' &&\equiv \mathfrak{u}(2,2|4) .
\end{alignat*}
As usual, the subspace of $\mathfrak{su}(2,2|4)$ absent in $\mathfrak{psu}(2,2|4)$ is the ideal generated by $\mathrm{i}\,\mathbb{I}_{4|4}$ (i.e. the identity supermatrix in $\mathfrak{gl}(4|4)$ scaled by the imaginary unit). On the other hand, the subspace of $\mathfrak{pu}(2,2|4)$ absent in $\mathfrak{psu}(2,2|4)$ is the one spanned by the diagonal matrix $\mathrm{i} \Sigma$ where
\begin{equation} \label{DefSigma}
    \Sigma
        \equiv \begin{pmatrix}
            +\mathbb{I}_4 & 0 \\ 0 & -\mathbb{I}_4
        \end{pmatrix} ,
\end{equation}
where $\mathbb{I}_4$ is the $4\times4$ identity matrix.

As discussed later, a special role is played by the $\mathbb{Z}_4$-grading
\begin{align*}
    \mathfrak{g} &= \mathfrak{g}_{\bar{0}} + \mathfrak{g}_{\bar{1}} + \mathfrak{g}_{\bar{2}} + \mathfrak{g}_{\bar{3}} , \\
    \mathfrak{h} &= \mathfrak{g}_{\bar{0}} ;
\end{align*}
this decomposition depends on the choice of a marked point in $\text{AdS}_5 \times \mathbb{S}^5$. Here $\mathfrak{g}_{\bar{0}}$ contains the isometries leaving that point invariant, while $\mathfrak{g}_{\bar{2}}$ contains the isometries which act like shifts in the vicinity of the marked point; the subspaces $\mathfrak{g}_{\bar{1}}$ and $\mathfrak{g}_{\bar{3}}$ contain the supersymmetries.

The supermanifold $\text{AdS}_5 \times \mathbb{S}^5$ can be realized as the coset\footnote{$G$ and $H$ are the connected Lie groups associated to $\mathfrak{g}$ and $\mathfrak{h}$, respectively.} $H\backslash G$ by the left multiplication, i.e. the set of $g\in G$ modulo the equivalence relation
\begin{equation} \label{Denominator}
    g\simeq hg , \qquad \text{for}\ h\in H .
\end{equation}
This is the gauge symmetry of the worldsheet sigma-model, and there exists a gauge choice in which
\begin{equation*}
    g = \exp\bigl(x^m t^2_m + \theta_{\text{L}}^{\alpha} \, t^3_{\alpha} + \theta_{\text{R}}^{\beta} \, t^1_{\beta}\bigr) ;
\end{equation*}
$x$ and $\theta$ are the coordinates on $\text{AdS}_5 \times \mathbb{S}^5$. Here $m$ runs on $\{1,\ldots,9\}$, labelling the translation generators $t^2_m$; in turn, $\alpha$ and $\beta$ run on $\{1,\ldots,16\}$; $t^3_\alpha$ and $t^1_\beta$ are the fermionic generators of left and right supersymmetries, respectively. The global symmetries correspond to constant right shifts of $g$.

For $\lambda_{\bar 3}\in\mathfrak{g}_{\bar 3}$ and $\lambda_{\bar 1}\in\mathfrak{g}_{\bar 1}$, we shall abbreviate
\begin{align}
    \lambda_{\bar 3}^2
        &:= \frac{1}{8} \operatorname{STr}\bigl( \Sigma \lambda_{\bar 3}^2 \bigr) , \label{AbbreviateC2L}
    \\
    \lambda_{\bar 1}^2
        &:= \frac{1}{8} \operatorname{STr}\bigl( \Sigma \lambda_{\bar 1}^2 \bigr) , \label{AbbreviateC2R}
\end{align}
where $\operatorname{STr}$ is the supertrace in the fundamental representation.

We will work at the level of Taylor series of functions on $\text{AdS}_5 \times \mathbb{S}^5$, and the complexified $\text{AdS}_5 \times \mathbb{S}^5$. Complexified formulas are often more streamlined. For example, the supersymmetry algebra becomes the projective special linear Lie superalgebra, and the stabilizer of a point becomes a direct sum of symplectic Lie algebras:
\begin{align*}
    \mathfrak{g} &= \mathfrak{psl}(4|4,\mathbb{C}) , \\
    \mathfrak{h} &= \mathfrak{sp}(2,\mathbb{C})\oplus \mathfrak{sp}(2,\mathbb{C}) .
\end{align*}

\subsubsection{Pure spinor bundle and BRST operator}\label{sec:PureSpinorBundle}

The pure spinor variables parametrizing the pure spinor cone in the spin bundle over $G/H$ are
\begin{align*}
    \lambda_{\text{L}} \in \mathfrak{g}_{\bar{3}}
    \qquad\text{and}\qquad
    \lambda_{\text{R}} \in \mathfrak{g}_{\bar{1}} ,
\end{align*}
satisfying
\begin{equation} \label{PureSpinorConditions}
    \{\lambda_{\text{L}},\lambda_{\text{L}}\} = \{\lambda_{\text{R}},\lambda_{\text{R}}\} = 0 .
\end{equation}
Let $\text{Cone}_{\text{L}} \times \text{Cone}_{\text{R}}$ denote the subspace in $\mathfrak{g}_{\bar{3}} \oplus \mathfrak{g}_{\bar{1}}$ defined by \eqref{PureSpinorConditions}; here $\text{Cone}_{\text{L}}$ and $\text{Cone}_{\text{R}}$ are the so-called ``pure spinor cones''.

We shall study vertex operators as formal Taylor series in a vicinity of a point in $\text{AdS}_5 \times \mathbb{S}^5$. Formally, ghost number $n$ vertices are elements of the coinduced representation
\begin{equation*}
    \operatorname{Coind}_{\mathfrak{h}}^{\mathfrak{g}} \mathcal{P}^n ,
\end{equation*}
where $\mathcal{P}^n$ is the space of polynomials of order $n$ on the pure spinor variables $\lambda_{\text{L}}$ and $\lambda_{\text{R}}$ --- i.e. polynomial functions on $\text{Cone}_{\text{L}} \times \text{Cone}_{\text{R}}$.
The space $\operatorname{Coind}_{\mathfrak{h}}^{\mathfrak{g}} \mathcal{P}^n$ is essentially the space of Taylor series of $x$, $\theta$ and $\lambda$ which are polynomial in $\lambda$.

The action of the BRST operator on a vertex operator $V$ is given by
\begin{equation} \label{BRSTOperator}
    QV(g,\lambda_{\text{L}},\lambda_{\text{R}})
        = \frac{d}{d\epsilon} \biggr|_{\epsilon=0} V\bigl( \mathrm{e}^{\epsilon \, (\lambda_{\text{L}}^{\alpha}t^3_{\alpha} + \lambda_{\text{R}}^{\alpha} t^1_{\alpha})} \, g,\lambda_{\text{L}},\lambda_{\text{R}} \bigr) .
\end{equation}
Geometrically, the pure spinor bundle over $\text{AdS}_5 \times \mathbb{S}^5$ is
\begin{equation} \label{PSAdS}
    PS(\text{AdS}_5 \times \mathbb{S}^5)
        = \frac{\text{Cone}_{\rm L}\times \text{Cone}_{\rm R} \times G}{H} .
\end{equation}

\subsubsection{Ten-dimensional gamma-matrices} \label{sec:TenDimensionalGammaMatrices}

The ten-dimensional gamma-matrices satisfy:
\begin{equation*}
  \Gamma^m_{\alpha\beta}
    = \Gamma^m_{\beta\alpha}
  \,,\quad
  \Gamma_m^{\alpha\beta}
    = \Gamma_m^{\beta\alpha}
  \,,\quad
  \Gamma_{(m}^{\alpha\rho} \Gamma_{n)\rho\beta}^{}
    = \delta_{mn}^{} \delta^{\alpha}_{\beta}
  \,,\quad
  \Gamma_{(m|\alpha\rho}^{} \Gamma_{|n)}^{\rho\beta}
    = \delta_{mn}^{} \delta_{\alpha}^{\beta} .
\end{equation*}
We will denote
\begin{align*}
    (\widehat{\lambda\theta})^{\alpha\beta}
        &= \lambda^{\alpha'} \Gamma_{\alpha'\beta'}^m \theta^{\beta'} \; \Gamma_m^{\alpha\beta} ,
    \\
    (\theta\psi)
        &= \theta^{\alpha}\psi_{\alpha} ,
    \\
        &\qquad \text{etc.}
\end{align*}

\subsection{Cohomology complex of \texorpdfstring{$\mathfrak{g}$}{frak(g)}} \label{sec:CohomologyComplex}

When we present the pure spinor bundle over $\text{AdS}_5 \times \mathbb{S}^5$ as in \eqref{PSAdS}, the subgroup $H$ acts on $G$ from the left, as in equation \eqref{Denominator} --- and by the conjugation on $\text{Cone}_{\text{L}} \times \text{Cone}_{\text{R}}$. The global \textbf{right} shifts of $G$ remain a symmetry of the construction.

In order to define the Lie superalgebra cohomological complex, we add extra coordinates which are essentially the Faddeev--Popov ghosts. Namely, we replace the pure spinor bundle in \eqref{PSAdS} by
\begin{equation} \label{SuperspaceM}
    M = \frac{\text{Cone}_{\text{L}} \times \text{Cone}_{\text{R}} \times \Pi TG}{H} .
\end{equation}
where $\Pi TG$ is the statistics-reversed tangent bundle of $G$ \cite{Alexandrov:1995kv}
The extra coordinates, parametrizing the fibre of $\Pi TG$ shall be denoted $C^a$.

The Lie superalgebra cohomological differential is the following odd nilpotent vector field on $M$:
\begin{equation} \label{DLieAsVectorField}
    d_{\rm Lie} V(g,\lambda_{\text{L}},\lambda_{\text{R}},C)
        = \left.\frac{d}{d\epsilon}\right|_{\epsilon=0} V\bigl( g \, \mathrm{e}^{\epsilon \, C^a t_a}, \lambda_{\text{L}}, \lambda_{\text{R}},C \bigr) .
\end{equation}
This is perhaps similar to (\ref{BRSTOperator}), except that now we do right shifts of $g$ instead of left shifts.

As a side remark, we could have also imposed the pure spinor constraints on $C$.
We are not sure if that would be useful or physically meaningful.
The main point of having ``normal'' unconstrained ghost $C$ is that the cohomology of $d_{\rm Lie}$
is much simpler (smaller) than the cohomology of $Q$.

\subsection{Shapiro's lemma}\label{ShapiroLemma}

In the vicinity of the unit, we can parametrize a point $g \in G$ as
\begin{equation*}
    g = \mathrm{e}^{\omega} \, \mathrm{e}^Z ,
    \qquad\text{where}\
    \omega \in \mathfrak{h}
    \ \text{and}\
    Z\in \mathfrak{h}^{\perp} .
\end{equation*}
Then functions on $M$ (defined in (\ref{SuperspaceM})) can be expressed through the following basic coordinates:
\begin{align*}
    C_{\parallel} = \mathrm{e}^{-\omega} \, d\mathrm{e}^{\omega} 
    &\colon M \longrightarrow \Pi \mathfrak{h} ,
    \\
    dZ &\colon M \longrightarrow \Pi (\mathfrak{g}/\mathfrak{h}) ,
    \\
    l^{\alpha} = \bigl( \mathrm{e}^{-\mathrm{ad}_\omega} \lambda \bigr)^\alpha
    &\colon M \longrightarrow \mathbb{C} .
\end{align*}
Consider the embedding of $\Pi TH/H$ into a formal neighborhood of $\Pi TH/H$ in $\Pi TG/H$, more precisely:
\begin{equation*}
    \frac{\text{Cone}_{\text{L}} \times \text{Cone}_{\text{R}} \times \Pi TH}{H} 
    \longrightarrow
    \frac{\text{Cone}_{\text{L}} \times \text{Cone}_{\text{R}} \times \Pi TG}{H} .
\end{equation*}
This induces the embedding of the space of functions of $C_{\parallel}$ and $l^{\alpha}$ into the space of functions of $C_{\parallel}$, $l^{\alpha}$, $Z$ and $dZ$, understood as formal Taylor series in $Z$ and $dZ$. Shapiro's lemma tells us that this is a homotopy retract. The proof is essentially the same as of the Poincar\'{e} lemma.
\begin{equation*}
  f(Z,dZ) - f(0,0) = \left[\,{\cal L}_E^{-1}\iota_E\,,\;d\,\right] f(Z,dZ)
\end{equation*}
where $E = Z\partial_Z$.

\section{BRST cohomology}\label{Preliminaries}

\subsection{Ghost number one vertex operators}\label{sec:GhostNumberOne}

Ghost number one vertices were constructed in \cite{Vallilo:2003nx}. Let us denote
\begin{equation*}
    \Lambda\langle\xi\rangle
        \equiv \operatorname{STr}\bigl(\xi g^{-1}(\lambda_{\text{L}} - \lambda_{\text{R}}) \, g \bigr) ;
\end{equation*}
the Taylor expansion of $\Lambda$ in powers of $x$ and $\theta$ defines a map
\begin{equation} \label{DefLambda}
    \Lambda \colon \mathfrak{g} \longrightarrow \operatorname{Coind}_{\mathfrak{h}}^{\mathfrak{g}} \mathcal{P}^1
\end{equation}
such that, for all $\xi\in\mathfrak{g}$, we have
\begin{equation*}
    Q\Lambda\langle\xi\rangle = 0 ,
\end{equation*}
and $\Lambda\langle\xi\rangle$ represents a nontrivial class in $H^1_Q$. This map $\Lambda$ is an isomorphism; the first BRST cohomology corresponds to symmetries,
\begin{equation*}
    H^1_Q = \mathfrak{g} .
\end{equation*}

\subsection{Ghost number two vertex operators} \label{sec:GhostNumberTwo}

\subsubsection{Dilaton zero mode and beta-deformation} \label{sec:DilatonAndBeta}

The simplest ghost number two vertex operator is the ``dilaton zero mode'' $\operatorname{Str}(\lambda_{\text{L}} \lambda_{\text{R}})$ \cite{Berkovits:2008ga}. This is actually a misnomer; it is better to call it ``density of RR 5-form field strenght'' $F_{ijklm}$. In the flat space limit, it can be gauge-fixed to take the form
\begin{align*}
    \bigl( \theta_{\text{L}} \, \widehat{\lambda_{\text{L}} \theta_{\text{L}}} \, \hat{F} \, \widehat{\theta_{\text{R}} \lambda_{\text{R}}}  \, \theta_{\text{R}} \bigr) ,
    \qquad\text{where}\
    \hat{F} = F_{ijklm} \Gamma^i \Gamma^j \Gamma^k \Gamma^l \Gamma^m ,
\end{align*}
and $\Gamma^i$ are the 10-dimensional Gamma matrices.

The next simplest vertex operator is the product of two ghost number one vertices. It transforms essentially in the wedge product of two adjoint representations \cite{Bedoya:2010qz},
\begin{equation} \label{BetaPrime}
    \beta'
        = \frac{\mathfrak{g}\wedge\mathfrak{g}}{\mathfrak{g}'} .
\end{equation}
The injection $\mathfrak{g}'\subset \mathfrak{g}\wedge\mathfrak{g}$ uses the ``split Casimir'' $C^{ab} \, t_a\otimes t_b$ given by
\begin{equation*}
    x \longmapsto C^{ab} \, t_a\wedge [t_b,x] .
\end{equation*}
However, only a subspace of $\beta'$ corresponds to actual infinitesimal deformations, namely
\begin{equation*}
    \beta =
        \frac{(\mathfrak{g}\wedge\mathfrak{g})_{0}}{\mathfrak{g}'}
        \simeq
        \frac{(\mathfrak{g}\wedge\mathfrak{g})_{\hat{0}}}{\mathfrak{g}} ;
\end{equation*}
here $(\mathfrak{g}\wedge\mathfrak{g})_0 \subset \mathfrak{g}\wedge\mathfrak{g}$ is the subspace with zero internal commutator, and $(\mathfrak{g}\wedge\mathfrak{g})_{\hat{0}} \subset \mathfrak{g}\wedge\mathfrak{g}$ is the subspace with zero internal commutator in $\widehat{\mathfrak{g}}$ (i.e. the internal commutator should be zero including the central element):
\begin{align*}
    \sum_i x_i\wedge y_i\in (\mathfrak{g}\wedge\mathfrak{g})_{\hat{0}}  
        \iff & \sum_i [x_i,y_i] = 0
        \quad\text{and}\quad \sum_i c_2(x_i,y_i) = 0 .
\end{align*}
The elements of $\beta'$ outside of $\beta$ are ``non-physical'' \cite{Bedoya:2010qz, Mikhailov:2012id, Arutyunov:2015mqj, Wulff:2016tju}, and the non-physical states suffer BRST anomaly. Non-physical beta-deformations are the only example of ghost number two vertices not corresponding to SUGRA solutions; all other ghost number two vertices correspond to some supergravity solutions, as we have
\begin{align*}
    H^2_{Q,\text{phys}} &\subset H^2_Q ,
    \\
    \frac{H^2_Q}{H^2_{Q,\text{phys}}}
        &= \frac{\beta'}{\beta}
        = \mathfrak{g} .
\end{align*}

\subsubsection{Vertices \emph{vs} SUGRA solutions} \label{sec:VerticesVsSUGRA}

On the other hand, not all linearized SUGRA solutions are captured by the vertices. There is an exact sequence:
\begin{equation} \label{Gh2VsSUGRA}
    0 \longrightarrow
        \mathbb{R}^2 \longrightarrow
        \mathcal{H}_{\text{SUGRA}} \longrightarrow 
        H^2_{Q,\text{phys}} \longrightarrow
        0 .
\end{equation}
The ${\mathbb R}^2 \subset {\cal H}_{\text{SUGRA}}$ is spanned by the constant dilaton and the constant axion; neither of those is ``visible'' as a ghost number two vertex. However, we will see that they are visible as ghost number three vertices.

\subsubsection{Rolling dilaton and axion}\label{sec:Rolling}

We will now describe two special solutions of linearized SUGRA equations. Recall that $\text{AdS}_5$ can be realized as a hyperboloid in $\mathbb{R}^{2+4}$. Let us introduce complex coordinates $Z_0,Z_1,Z_2$ in $\mathbb{R}^{2+4}$,
so that the flat metric is given by $|dZ_0|^2 - |dZ_1|^2 - |dZ_2|^2$. The coordinate $T$, defined as
\begin{equation*}
    T = \frac{1}{\mathrm{i}} \log(Z/\overline{Z}) ,
\end{equation*}
can be understood as a ``global time''.

Consider now the following two solutions of the classical SUGRA equations for dilaton $\phi$ and axion $a$:
\begin{align}
     \phi   &= \alpha \, T , \label{RollingDilaton} \\
     a      &= \alpha \, T , \label{RollingAxion}
\end{align}
for a constant coefficient $\alpha$. We do not know the explicit formulas for the corresponding ghost number two vertices, so this is an interesting open problem --- but we do know that they exist. Let us denote them by $V_{\text{linear} \atop \text{dilaton}}$ and $V_{\text{linear} \atop \text{axion}}$.

\subsection{Ghost number three vertex operators}\label{sec:GhostNumberThree}

Ghost number three vertex operators
are \emph{almost} in one-to-one correspondence with linearized SUGRA solutions \cite{Mikhailov:2014qka},
\emph{except} that they miss the constant RR flux in $\mathbb{S}^5$. That constant RR flux corresponds to $\operatorname{Str}(\lambda_{\text{L}}\lambda_{\text{R}})$
in the ghost number two, but there is no corresponding vertex at the ghost number three. There are no non-physical vertices at the ghost number three; all ghost number three vertices correspond to SUGRA solutions.

Ghost number three vertices can be obtained by multiplying a ghost number two vertex with a ghost number one vertex. For any ghost number two vertex $V$, and any $\xi\in\mathfrak{g}$, the product $\Lambda\langle\xi\rangle \, V$ is a ghost number three vertex. Let us restrict ourselves to physical $V$ only.
Consider the map $[[-]]$ from ghost number three vertices of the form $\Lambda\langle\xi\rangle\,V$ to ghost number two vertices:
\begin{equation} \label{InnerBracket}    
    \bigl[\!\bigl[ \Lambda\langle\xi\rangle \, V \bigr]\!\bigr] = \xi \mathbin{.} V ,
\end{equation}
where $\xi \mathbin{.} V$ is the action of $\xi$ on $V$.

This is a surjective map in cohomology. It is \emph{almost} an isomorphism between ghost number three vertices and ghost number two vertices, \emph{except} that it has a two-dimensional kernel. Its kernel is spanned by the elements
\begin{align}
    W_{\text{dilaton} \atop \text{0-mode}}
        &= \Lambda\left\langle \frac{\partial}{\partial T}\right\rangle \; V_{\text{linear} \atop \text{dilaton}} ,
        \label{Gh3DilatonZeroMode}
    \\
    W_{\text{axion} \atop \text{0-mode}}
        &= \Lambda\left\langle\frac{\partial}{\partial T}\right\rangle \; V_{\text{linear} \atop \text{axion}} .
        \label{Gh3AxionZeroMode}
\end{align}
These elements represent BRST-nontrivial elements of ghost number three cohomology, but Eq. \eqref{InnerBracket} gives zero, i.e.
\begin{equation*}
    \left[\!\left[
        \Lambda\left\langle\frac{\partial}{\partial T}\right\rangle \; V_{\text{linear} \atop \text{dilaton}}
    \right]\!\right]
\quad\text{and}\quad
    \left[\!\left[
        \Lambda\left\langle\frac{\partial}{\partial T}\right\rangle \; V_{\text{linear} \atop \text{axion}}
    \right]\!\right]
    \quad\text{are BRST-trivial} .
\end{equation*}
Therefore, there are the following exemptions from the rule that $[[-]]$ is an isomorphism between
ghost number three and ghost number two: one is that $W_{\text{dilaton} \atop \text{0-mode}}$ and $W_{\text{axion} \atop \text{0-mode}}$ go to zero; the other is that there is nothing corresponding to $\operatorname{Str}(\lambda_{\text{L}}\lambda_{\text{R}})$ in ghost number three.

Of course, $W_{\text{dilaton} \atop \text{0-mode}}$ and $W_{\text{axion} \atop \text{0-mode}}$ are not covariant; the cohomology is $\mathfrak{g}$-invariant, but the vertices themselves are not. However, we will show in
Section \ref{sec:TwoZeroModesGh3} that $W_{\text{axion} \atop \text{0-mode}}$ can at least be chosen to transform
in a finite-dimensional representation of $\mathfrak{g}$. We are not sure if this is also true for $W_{\text{dilaton} \atop \text{0-mode}}$. The cohomological obstacle to the covariance of these vertices is a cohomology group with
values in a finite dimensional representation. It is obtained by acting on them with higher differentials (see Section \ref{KnightMove}). It would be interesting to find explicit expressions for $V_{\text{linear} \atop \text{dilaton}}$, $V_{\text{linear} \atop \text{axion}}$,
$W_{\text{dilaton} \atop \text{0-mode}}$, $W_{\text{axion} \atop \text{0-mode}}$.

Ghost number three cohomology misses the vertex corresponding to $\operatorname{Str}(\lambda_{\text{L}}\lambda_{\text{R}})$. This is related to the existence of an invariant subspace in $H_Q^2$, complementary to the one-dimensional subspace generated by $\operatorname{Str}(\lambda_{\text{L}}\lambda_{\text{R}})$. Hence, the relation to a linearized SUGRA solutions is
\begin{equation} \label{Gh3VsSUGRA}
    {\cal H}_{\text{SUGRA}} = \mathbb{R} \oplus H_Q^3 ,
\end{equation}
where the subspace $\mathbb{R}$ is generated by $\operatorname{Str}(\lambda_{\text{L}}\lambda_{\text{R}})$, i.e. the RR 5-form flux in the $\mathbb{S}^5$ direction. This looks similar to Eq. \eqref{Gh2VsSUGRA}, except that the short exact sequence in \eqref{Gh2VsSUGRA} does not split.

In linearized SUGRA around $\text{AdS}_5 \times \mathbb{S}^5$ background, the constant RR flux in $\mathbb{S}^5$ is a SUSY-invariant solution, and the one-dimensional subspace generated by it has an invariant complementary subspace. (The complementary subspace is the space of those solutions which are realized as ghost number three vertex operators.) This is different from the dilaton and axion zero modes, which are invariant but do not split out.

Ghost number three operators may be useful for the computation of string amplitudes,
see \cite{Kashyap:2025rvf} for an application to the two-point amplitudes.

\subsection[{Class \texorpdfstring{$\operatorname{Str}(\lambda_{\text{L}}\lambda_{\text{R}})$}{STr(λL λR)} is a discrete state}]{Class \texorpdfstring{$\boldsymbol{\operatorname{Str}(\lambda_{\text{L}}\lambda_{\text{R}})}$}{STr(λL λR)} is a discrete state} \label{sec:DiscreteState}

The splitting out of the zero mode corresponding to $\operatorname{Str}(\lambda_{\text{L}}\lambda_{\text{R}})$ is not accidental. As we will now explain, this zero mode corresponds to the number of colours of the dual gauge theory:
\begin{equation*}
    \int_{\mathbb{S}^5} F_5 = N .
\end{equation*}
If this zero mode were not split out, there would exist a solution with ``rolling'' $F_5$ analogous to those given by \eqref{RollingDilaton} and \eqref{RollingAxion}. That would contradict $N$ being an integer.

The worldsheet pure spinor sigma-model in $\text{AdS}_5 \times \mathbb{S}^5$ is:
\begin{align}
    S &= (R_{\text{AdS}})^2 \int d^2 z \; \operatorname{Str} \left( \frac{1}{2} J_{2+} J_{2-} + \ldots \right) ,
    \nonumber \\
    (R_{\text{AdS}})^2 &= \sqrt{g_{\text{s}} N} , \label{RvsGN}
\end{align}
where $R_{\text{AdS}}$ is the radius of $\mathrm{AdS}_5$, and $g_{\text{s}} = g_{\text{YM}}^2$ is the string coupling constant, which is the dilaton expectation value, $g_{\text{s}} = \mathrm{e}^{\phi}$. The dilaton enters into the worldsheet sigma-model through the coupling to the worldsheet curvature (the Fradkin--Tseytlin term),
\begin{equation} \label{FradkinTseytlinTerm}
    \frac{1}{4\pi} \int d^2 z \; \phi \sqrt{h} \, \mathcal{R} ,
\end{equation}
where $\mathcal{R}$ is the worldsheet Ricci scalar.

When we insert the integrated vertex operator $U = \operatorname{Str}\left(\frac{1}{2} J_{2+} J_{2-} + \ldots\right)$, we do not change the coupling to the worlsdsheet curvature given by \eqref{FradkinTseytlinTerm}. This means that $U$ corresponds to an infinitesimal variation of $N$ keeping $g_{\text{s}}$ fixed.

In the near-flat-space expansion, we have \cite{Mikhailov:2012id}
\begin{align*}
    S_{\text{AdS}}
        &= S_{\text{flat}}
        + \frac{1}{R_{\text{AdS}}} \int d^2z \; \bigl( \eta^{\alpha\hat{\alpha}} d_{\alpha} \hat{d}_{\hat{\alpha}}
        + \text{terms with $\theta$} \bigr)
    + O\left(\frac{1}{(R_{\text{AdS}})^2}\right)
    \\
    &= S_{\text{flat}} + \frac{1}{R_{\text{AdS}}} \int d^2z\; \eta^{\alpha\hat{\alpha}} S_{\alpha} \hat{S}_{\hat{\alpha}} + O\left(\frac{1}{(R_{\text{AdS}})^2}\right) .
\end{align*}
Notice that $\frac{1}{R_{\text{AdS}}} = \frac{Ng_{\text{s}}}{(R_{\text{AdS}})^5}$, and $\frac{N}{(R_{\text{AdS}})^5} = F_5$ is the density of the RR 5-form. Therefore the integrated vertex
is of the form \cite{Berkovits:1999zq}
\begin{equation*}
    g_{\text{s}} \, F_5^{\alpha\hat{\alpha}} \, d_{\alpha} \hat{d}_{\hat{\alpha}} .
\end{equation*}

\section{AdS as a deformation of flat space}\label{FlatSpace}

Our notations for gamma-matrices are summarized in Section \ref{sec:TenDimensionalGammaMatrices}.

\subsection{AdS superalgebra using flat-space notations}\label{sec:FlatSpacePrelims}

In a vicinity of a point, $\text{AdS}_5 \times \mathbb{S}^5$ can be constructed as a deformation of flat space; then, it is natural to use Gamma-matrix notations for the generators of $\mathfrak{g}$. The (anti-)commutators of supersymmetries are
\begin{align*}
    \bigl\{s^{\text{L}}_{\alpha} , s^{\text{L}}_{\beta} \bigr\} 
        &= \Gamma_m^{\alpha\beta} P_m^{} ,
    \\
    \bigl\{ s^{\text{R}}_{\hat{\alpha}} , s^{\text{R}}_{\hat{\beta}} \bigr\}
        &= \Gamma^m_{\hat{\alpha}\hat{\beta}} P_m^{} ,
    \\
    \bigl[ P_m , s^{\text{L}}_{\alpha} \bigr]
        &= {}_{\alpha}(\Gamma_m\hat{F})^{\hat{\beta}} \, s^{\text{R}}_{\hat{\beta}} ,
    \\
    \bigl[ P_m , s^{\text{R}}_{\hat{\alpha}} \bigr]
        &= - {}_{\hat{\alpha}}(\Gamma_m\hat{F})^{\beta} \, s^{\text{L}}_{\beta} .
\end{align*}
In particular,
\begin{align*}
    \bigl[ P_m^{} , \{ s^{\text{L}}_{\alpha} , s^{\text{R}}_{\hat{\beta}} \} \bigr]
        &= {}_{\alpha}\bigl(\Gamma_m \hat{F} \bigr){}^{\hat{\alpha}}
        \, \bigl\{ s^{\text{R}}_{\hat{\alpha}} , s^{\text{R}}_{\hat{\beta}} \bigr\}
        -
        {}_{\hat{\beta}}\bigl(\Gamma_m \hat{F} \bigr){}^{\beta}
        \, \bigl\{ s^{\text{L}}_{\alpha}, s^{\text{L}}_{\beta} \bigr\}
        \\
        &= {}_{\alpha}\bigl(\Gamma_m \hat{F} \Gamma_n\bigr)_{\hat{\beta}} \, P_n^{}
        -
        {}_{\hat{\beta}} \bigl(\Gamma_m \hat{F} \Gamma_n\bigr)_{\alpha} \, P_n^{}
        \\
        &= {}_{\alpha}\bigl( (\Gamma_m \overline{\Gamma}_n - \Gamma_n \overline{\Gamma}_m) \hat{F} \bigr)_{\hat{\beta}} \, P_n^{} .
\end{align*}
The supertrace in the fundamental representation corresponds to:
\begin{equation*}
    \operatorname{Str}(s^{\text{L}}_{\alpha} s^{\text{R}}_{\beta})
        = -\operatorname{Str}(s^{\text{R}}_{\beta} s^{\text{L}}_{\alpha})
        = \hat{F}^{-1}_{\alpha\beta} .
\end{equation*}

\subsection{Outer derivation}\label{sec:OuterDer}

Recalling \eqref{DefSigma}, we have
\begin{align*}
    \bigl[ \Sigma , s^{\text{L}}_{\alpha} \bigr]
        &= +\mathrm{i} \, s^{\text{R}}_{\alpha} ,
    \\
    \bigl[ \Sigma , s^{\text{R}}_{\alpha} \bigr]
        &= -\mathrm{i} \, s^{\text{L}}_{\alpha} .
\end{align*}

\subsection{Worldsheet parity symmetry}\label{sec:WorldsheetParity}

Let $P_{\text{w-s}}$ denote the wordsheet parity transformations
\begin{align*}
    P_{\text{w-s}} \colon
    \begin{cases}
        s^{\text{L}}_{\alpha}
            \longmapsto s^{\text{R}}_{\alpha} ,
        \\
        s^{\text{R}}_{\alpha}
            \longmapsto s^{\text{L}}_{\alpha} ,
        \\
        F   \longmapsto -F .
    \end{cases}
\end{align*}
The formalism is invariant under the parity transformations, provided that we also change the sign
of the RR 5-form $F$.\footnote{
    Notice that $( (\Gamma_m \overline{\Gamma}_n - \Gamma_n \overline{\Gamma}_m) \, F \bigr)_{\alpha\beta}$ is antisymmetric in $\alpha\leftrightarrow\beta$.
}

Notice that $\hat{F}^{-1}_{\alpha\beta} = \hat{F}^{-1}_{\beta\alpha}$, thus the 5-form flux zero mode
\begin{equation*}
    \operatorname{Str}(\lambda_{\text{L}}\lambda_{\text{R}})
        = \lambda_{\text{L}}^{\alpha} \, F^{-1}_{\alpha\beta} \, \lambda_{\text{R}}^{\beta}
\end{equation*}
is parity-odd:
\begin{equation*}
    \lambda_{\text{L}}^{\alpha} \, F^{-1}_{\alpha\beta} \, \lambda_{\text{R}}^{\beta}
    \quad
    \underset{
        \left(\substack{
        \lambda_{\text{L}}\leftrightarrow\lambda_{\text{R}} \\
        F\to -F}\right)
    }{\longrightarrow}
    \quad
    - \lambda_{\text{L}}^{\alpha} \, F^{-1}_{\alpha\beta} \, \lambda_{\text{R}}^{\beta} .
\end{equation*}

The central charge is of $\mathfrak{g}$ is given by
\begin{align}
    c_2 &= \operatorname{Str}
    \left(
      \bigl( \gamma_{\text{L}}^{\alpha} \, s^{\text{L}}_{\alpha} + \gamma_{\text{R}}^{\alpha} \, s^{\text{R}}_{\alpha} \bigr)
      \bigl[ \Sigma, \bigl(\gamma_{\text{L}}^{\beta} \, s^{\text{L}}_{\beta} + \gamma_{\text{R}}^{\beta} \, s^{\text{R}}_{\beta} \bigr) \bigr]
      \right)
    \nonumber \\
        &= \operatorname{Str}
    \left(
      \bigl( \gamma_{\text{L}}^{\alpha} \, s^{\text{L}}_{\alpha} + \gamma_{\text{R}}^{\alpha} \, s^{\text{R}}_{\alpha} \bigr)
      \bigl( \mathrm{i} \gamma_{\text{L}}^{\beta} \, s^{\text{R}}_{\beta} - \mathrm{i} \gamma_{\text{R}}^{\beta} \, s^{\text{L}}_{\beta} \bigr)
      \right)
    \nonumber \\
    &= \mathrm{i} \gamma_{\text{L}}^{\alpha} \, F^{-1}_{\alpha\beta} \, \gamma_{\text{L}}^{\beta}
    + \mathrm{i} \gamma_{\text{R}}^{\alpha} \, F^{-1}_{\alpha\beta} \, \gamma_{\text{R}}^{\beta} .\label{CentralChargeInFlatSpaceNotations}
\end{align}
We observe that $c_2$, as a function of $\gamma_{\text{L}}$, $\gamma_{\text{R}}$ and $F$, is odd under the transformation
\begin{equation*}
    \bigl(\gamma_{\text{L}}^{\alpha} , \gamma_{\text{R}}^{\alpha} , F^{-1}_{\alpha\beta} \bigr)
    \overset{P_{\text{w-s}}}{\longmapsto}
    \bigl( \gamma_{\text{R}}^{\alpha} , \gamma_{\text{L}}^{\alpha} , -F^{-1}_{\alpha\beta} \bigr) ;
\end{equation*}
thus, we conclude that the central charge $c_2$ is parity-odd.

\subsection{Zero modes at ghost number three}\label{sec:TwoZeroModesGh3}

Let us start by classifying Lorentz-invariant BRST-nontrivial expressions of the form $\lambda^3 \, \theta^5$; there are three candidates, namely
\begin{gather*}
    \bigl( \theta_{\text{L}} \, \widehat{\lambda_{\text{L}} \, \theta_{\text{L}}}^3 \, \theta_{\text{L}} \bigr) ,
    \\
    \bigl( \theta_{\text{R}} \, \widehat{\lambda_{\text{R}} \, \theta_{\text{R}}}^3 \, \theta_{\text{R}} \bigr) ,
    \\
    \left( \theta_{\text{L}} \, \widehat{\lambda_{\text{L}}\theta_{\text{L}}} \,
        \bigl( \widehat{\lambda_{\text{L}} \theta_{\text{L}}} - \widehat{\lambda_{\text{R}} \, \theta_{\text{R}}} \bigr)
        \widehat{\lambda_{\text{R}} \theta_{\text{R}}} \, \theta_{\text{R}} \right) .
\end{gather*}
However, one of these combinations is BRST-exact. Indeed, note that
\begin{align*}
    Q\Bigl( &
        x^2 \, \theta_{\text{L}} \Gamma^m \lambda_{\text{L}} \; \theta_{\text{R}} \Gamma^m \lambda_{\text{R}}
        \,+\,
        x^p \bigl( (\theta_{\text{L}})^3 \lambda_{\text{L}} \bigr)^{[pm]} \theta_{\text{R}} \Gamma^m \lambda_{\text{R}} + {}
    \\
    &\quad {} +
        x^p \bigl( (\theta_{\text{R}})^3 \lambda_{\text{R}} \bigr)^{[pm]} \theta_{\text{L}} \Gamma^m \lambda_{\text{L}}
        \,+\,
        \bigl[ (\theta_{\text{R}})^3 \lambda_{\text{R}} \bigr]^{[pm]} \bigl[ (\theta_{\text{L}})^3 \lambda_{\text{L}} \bigr]^{[pm]} \Bigr)
    \\
    &= \bigl( \theta_{\text{L}} \Gamma^p \lambda_{\text{L}} \bigr) \, \bigl( (\theta_{\text{L}})^3 \lambda_{\text{L}} \bigr)^{[pm]} \, \theta_{\text{R}} \Gamma^m \lambda_{\text{R}}
    - (\text{L}\leftrightarrow \text{R}) ;
\end{align*}
this is BRST-equivalent to the parity-odd expression
\begin{equation} \label{GhostNumber3Exact}
    \bigl( \theta_{\text{L}} \, \widehat{\lambda_{\text{L}} \theta_{\text{L}}}^3 \, \theta_{\text{L}} \bigr) - (\text{L}\leftrightarrow\text{R}) .
\end{equation}
While the parity-odd expression in \eqref{GhostNumber3Exact} above is exact, the parity-even combination
\begin{equation} \label{GhostNumber3Even}
    \Xi_{\text{L}\leftrightarrow\text{R} \; \text{even}}
        = \bigl( \theta_{\text{L}} \, \widehat{\lambda_{\text{L}} \theta_{\text{L}}}^3 \, \theta_{\text{L}} \bigr) + (\text{L}\leftrightarrow \text{R})
\end{equation}
is nontrivial and can be completed to a BRST-closed expression in $\text{AdS}_5 \times \mathbb{S}^5$.

In flat space, there exists the second nontrivial expression of the type $\lambda^3 \, \theta^5$, namely
\begin{align} \label{GhostNumber3Odd}
    \Xi_{\text{L}\leftrightarrow\text{R} \; \text{odd}} 
    = \bigl( \theta_{\text{L}} \, \widehat{\theta_{\text{L}} \lambda_{\text{L}}}^2 \, \widehat{\theta_{\text{R}} \lambda_{\text{R}}} \; \theta_{\text{R}} \bigr) . 
\end{align}
In Section 7 of \cite{Mikhailov:2014qka}, $\Xi_{\text{L} \leftrightarrow\text{R} \; \text{odd}}$ is an element of ${\cal E}^{1,2}$,  which ``corresponds to $R$ being constant times the unit matrix'', while $\Xi_{\text{L}\leftrightarrow\text{R} \; \text{even}}$ corresponds to an element of ${\cal E}^{2,1}$ which ``does not seem to cancel with anything''
(we do not have a rigorous proof that it is not exact).

The action of $\delta Q \equiv Q_{\text{AdS}} - Q_{\text{flat}}$ on both $\Xi_{\text{L}\leftrightarrow\text{R} \; \text{even}}$ and $\Xi_{\text{L}\leftrightarrow\text{R} \; \text{odd}}$ results in an expression of the order $\lambda^4 \, \theta^6$. There are the following types of expressions which cannot be resolved without $x$:
\begin{gather*}
    \Bigl( \theta_{\text{L}} \, \widehat{\theta_{\text{L}} \lambda_{\text{L}}}^3 \, \theta_{\text{L}} \Bigr) \bigl( \theta_{\text{R}} \lambda_{\text{R}} \bigr)^m ,
    \\
    \Bigl( \theta_{\text{R}} \, \widehat{\theta_{\text{R}} \lambda_{\text{R}}}^3 \, \theta_{\text{R}} \Bigr) \bigl( \theta_{\text{L}} \lambda_{\text{L}} \bigr)^m ,
    \\
    \bigl( \widehat{\theta_{\text{L}} \lambda_{\text{L}}}^2 \theta_{\text{L}} \bigr)^{\alpha} \bigl( \widehat{\theta_{\text{R}} \lambda_{\text{R}}}^2 \theta_{\text{R}} \bigr)^{\hat{\alpha}} .
\end{gather*}
None of them combines with $F^{\alpha\hat{\alpha}}$ into a Lorentz-invariant expression, thus they are absent. Therefore, there exists $\Xi_{\text{L}\leftrightarrow\text{R} \; \text{even}}'$ and $\Xi_{\text{L}\leftrightarrow\text{R} \; \text{odd}}'$ of the order $\lambda^4 \, \theta^8$, not containing
$x$, such that
\begin{align*}
    Q_{\text{AdS}}\bigl(
        \Xi_{\text{L}\leftrightarrow\text{R} \; \text{even}} + \Xi_{\text{L}\leftrightarrow\text{R} \; \text{even}}' \bigr)
        &= (\lambda^4 \, \theta^8 )_{\text{no-}x} , \\
    Q_{\text{AdS}}\bigl(
        \Xi_{\text{L}\leftrightarrow\text{R} \; \text{odd}} + \Xi_{\text{L}\leftrightarrow\text{R} \; \text{odd}}' \bigr)
        &= (\lambda^4 \, \theta^8)_{\text{no-}x} .
\end{align*}

In flat space, expressions of the type $\lambda^4 \, \theta^8$ which cannot be gauged away by $x$-independent terms are of the type
\begin{align*}
    T^{mn} = \bigl( \theta_{\text{L}} \, \widehat{\theta_{\text{L}} \lambda_{\text{L}}} \, \Gamma^m \, \widehat{\theta_{\text{L}} \lambda_{\text{L}}} \, \theta_{\text{L}} \bigr) \bigl( \theta_{\text{R}} \, \widehat{\theta_{\text{R}} \, \lambda_{\text{R}}} \, \Gamma^n \, \widehat{\theta_{\text{R}} \lambda_{\text{R}}} \, \theta_{\text{R}} \bigr) .
\end{align*}
This should be multiplied by $F^{\alpha\hat{\beta}} F^{\gamma\hat{\delta}}$, and the only invariant is
\begin{equation} \label{L4T8}
    \bigl( \theta_{\text{L}} \, \widehat{\theta_{\text{L}}\lambda_{\text{L}}} \, \widehat{\theta_{\text{L}}\lambda_{\text{L}}} \, \theta_{\text{L}} \bigr)^m
    \;
    \bigl( \theta_{\text{R}} \, \widehat{\theta_{\text{R}}\lambda_{\text{R}}} \, \widehat{\theta_{\text{R}}\lambda_{\text{R}}} \, \theta_{\text{R}} \bigr)^n
    \;
    F^{\alpha\hat{\beta}} \, \Gamma^m_{\alpha\gamma} \, \Gamma^n_{\hat{\beta}\hat{\delta}} \, F^{\gamma\hat{\delta}} .
\end{equation}
(Notice that the term $F^{\alpha\hat{\beta}} \, \Gamma^m_{\alpha\gamma} \, \Gamma^n_{\hat{\beta}\hat{\delta}} \, F^{\gamma\hat{\delta}}$ is the flat metric in the $\text{AdS}_5$ direction minus the flat metric in the $\mathbb{S}^5$ direction.)

Equation \eqref{L4T8} is the only obstacle to gauge away the terms $\lambda^4 \, \theta^8$. It is parity-even, thus there is no obstacle to completing $\Xi_{\text{L}\leftrightarrow\text{R} \; \text{odd}}$. In other words, there exists
\begin{equation*}
    W_{\text{axion} \atop \text{0-mode}}
        = \Xi_{\text{L}\leftrightarrow\text{R} \; \text{odd}} + \Xi_{\text{L}\leftrightarrow\text{R} \; \text{odd}}' + \Xi_{\text{L}\leftrightarrow\text{R} \; \text{odd}}'' + \ldots ,
\end{equation*}
not containing $x$, such that
\begin{equation*}
    Q_{\text{AdS}} \, W_{\text{axion} \atop \text{0-mode}} = 0 .
\end{equation*}
The terms $\Xi_{\text{L}\leftrightarrow\text{R} \; \text{odd}}'$, $\Xi_{\text{L}\leftrightarrow\text{R} \; \text{odd}}''$, $\ldots$ are of the order $\lambda^3 \, \theta^7$, $\lambda^3 \, \theta^9$, $\ldots$ respectively. The ghost number three vertex $W_{\text{axion} \atop \text{0-mode}}$ represents a cohomology class in $\text{AdS}_5 \times \mathbb{S}^5$ invariant under all even symmetries; in other words, $W_{\text{axion} \atop \text{0-mode}}$ is a section of the pure spinor bundle on $(\text{AdS}_5 \times \mathbb{S}^5)/G_{\text{even}}$.

The cohomology class of $W_{\text{axion} \atop \text{0-mode}}$ is $\mathfrak{g}$-invariant. Although $W_{\text{axion} \atop \text{0-mode}}$ itself is only $\mathfrak{g}$-invariant up to a $Q$-exact expression, we have just shown that it can be chosen to transform in a finite-dimensional representation (sections of the pure spinor bundle on $(\text{AdS}_5 \times \mathbb{S}^5)/G_{\text{even}}$). We do not have such an argument for the parity-even expression of Eq. (\ref{GhostNumber3Even}). However our arguments in Section \ref{sec:GhostNumberThree} imply that Eq. (\ref{GhostNumber3Even}) should also complete to an expression which is $\mathfrak{g}$-invariant up to $Q$-exact terms --- but we are not sure if the representative itself can be chosen in a finite-dimensional representation.

\section{Matrix realization of \texorpdfstring{$\mathfrak{g}$}{frak(g)}} \label{MatrixRealization}

The Lie superalgebra $\mathfrak{g}$ is a real form of $\mathfrak{psl}(4|4,\mathbb{C})$ --- or, in other words, $\mathfrak{psl}(4|4,\mathbb{C})$ is a complexification of $\mathfrak{g}$. The complexification of the stabilizer of a point of $\text{AdS}_5 \times \mathbb{S}^5$ is, as before,
\begin{equation*}
    \mathfrak{sp}(2,\mathbb{C}) \oplus \mathfrak{sp}(2,\mathbb{C}) \subset \mathfrak{g}_{\mathbb{C}} .
\end{equation*}
Consider $(4|4)\times (4|4)$ matrices
\begin{equation} \label{matrixdecompositionPSL}
    \left(
    \begin{array}{cc}
      A^{\alpha}_{\beta} & A^{\alpha}_b \\
      A^a_{\beta} & A^a_b
    \end{array}
    \right) ;
\end{equation}
each $\mathfrak{sp}(2,\mathbb{C})$ Lie subalgebras preserves a constant symplectic form, namely $\omega^{\alpha\beta}$ and $\omega_{ab}$. Following \cite{Mikhailov:2011af}, we use the notations
\begin{align*}
    X \cup Y
        &\equiv X^\alpha \omega_{\alpha\beta} Y^\beta ,
    \\
    X\cap Y
        &\equiv X_{\alpha} \omega^{\alpha\beta} Y_{\beta} ,
    \\
    (X\cap)^{\alpha}
        &\equiv X_{\alpha'}\omega^{\alpha'\alpha} .
\end{align*}
The left and right supersymmetries have $\mathbb{Z}_4$-grading $\bar{1}$ and $\bar{3}$, respectively. In matrix notations,
\begin{equation} \label{MatrixNotations}
    \gamma_{\text{L}}^{\alpha} \, s^{\text{L}}_{\alpha} + \gamma_{\text{R}}^{\alpha} \, s^{\text{R}}_{\alpha}
    =
    \left(
    \begin{array}{cc}
      0 & \gamma_{\text{L}} - \mathrm{i} \gamma_{\text{R}}
      \\
      \cap (\gamma_{\text{L}} + \mathrm{i} \gamma_{\text{R}}) \cup & 0
    \end{array}
    \right)
\end{equation}
and
\begin{equation*}
    \hat{F}{}^a_{\alpha}{}^b_{\beta} = \omega^{ab}\omega_{\alpha\beta} .
\end{equation*}

Finally, the parity transformation $\gamma_{\text{L}} \leftrightarrow \gamma_{\text{R}}$ acts as
\begin{equation}\label{ParityOnMatrix}
    \left(
    \begin{array}{cc}
      0 & \gamma_{\text{L}} - \mathrm{i} \gamma_{\text{R}} 
      \\
      \cap (\gamma_{\text{L}} + \mathrm{i} \gamma_{\text{R}}) \cup & 0
    \end{array}
    \right)
    \longmapsto
    \left(
    \begin{array}{cc}
      0 & -\mathrm{i}( \gamma_{\text{L}} + \mathrm{i} \gamma_{\text{R}}) 
      \\
      \cap \mathrm{i}(\gamma_{\text{L}} - \mathrm{i} \gamma_{\text{R}}) \cup & 0
    \end{array}
    \right) .
\end{equation}

\section{Symmetry bicomplex}\label{SymmetryBicomplex}

We will now review the construction presented in \cite{Mikhailov:2009rx, Mikhailov:2024sef}. The BRST complex of Type IIB supergravity in $\text{AdS}_5 \times \mathbb{S}^5$ is the coinduced representation
\begin{equation*}
    C_Q^p = \operatorname{Coind}_{\mathfrak{h}}^{\mathfrak{g}} \mathcal{P}^p ,
\end{equation*}
where $\mathcal{P}^p$ is the space of homogeneous $p$-th order polynomials on the pure spinors $\lambda_{\text{L}}$ and $\lambda_{\text{R}}$.

For any representation $\cal R$ of $\mathfrak{g}$, consider the double complex
\begin{equation*}
    C_\mathcal{R}^{p,q}
        = C^p \Bigl( \mathfrak{g} , \operatorname{Hom}\bigl( \mathcal{R} ,  \operatorname{Coind}_{\mathfrak{h}}^{\mathfrak{g}} \mathcal{P}^q \bigr) \Bigr)
\end{equation*}
with the differential
\begin{equation*}
    d_{\text{tot}} = d_{\text{Lie}} + Q .
\end{equation*}
As mentioned in Section \ref{Introduction}, there are two spectral sequences, whose second pages are
\begin{align*}
    E_2^{p,q}
        &= \operatorname{Ext}^p \bigl( \mathcal{R} , H^q(Q) \bigr) ,
    \\
    \widetilde{E}_2^{p,q}
        &= H_Q^p \Bigl( \operatorname{Ext}^q_{\mathfrak{g}} \bigl( \mathcal{R} , \operatorname{Coind}_{\mathfrak{h}}^{\mathfrak{g}} \mathcal{P}^{\bullet} \bigr)
      \Bigr) ,
\end{align*}
and both of them converging to the cohomology of $d_{\text{tot}}$:
\begin{equation*}
    \bigoplus_{p+q=n} \widetilde{E}_{\infty}^{p,q}
    = \bigoplus_{p+q=n} E_{\infty}^{p,q} .
\end{equation*}
The spectral sequence $E$ carries information about the pure spinor cohomologies; it tells us how the BRST cohomology is organized into representations of $\mathfrak{g}$. On the other hand, the spectral sequence $\widetilde{E}$ is much easier to work with, because
of Shapiro's lemma discussed in Section \ref{ShapiroLemma}.

\section{Lie superalgebra cohomology of \texorpdfstring{$\mathfrak{g}$}{frak(g)}} \label{CohomologyOfG}

Let $C$ denote the Faddeev--Popov ghosts of $\mathfrak{g}$.
Among them, we denote by $c^m$ the ghosts corresponding to translations, by $c^{[mn]}$ the ghosts corresponding to $\mathfrak{h}$, and by $\gamma$ the ghosts corresponding to $\mathfrak{g}_{\text{odd}}$. As a linear space, $\mathfrak{g}_{\text{odd}}$ is the direct sum of the upper-right and lower-left blocks:
\begin{equation*}
    \mathfrak{g}_{\text{odd}}
        = \mathfrak{g}_\uparrow \oplus \mathfrak{g}_\downarrow .
\end{equation*}
The Faddeev--Popov ghosts corresponding to $\mathfrak{g}_\uparrow$ will be denoted $\gamma^\uparrow$, and those corresponding to $\mathfrak{g}_\downarrow$ will be denoted $\gamma^\downarrow$. Thus both $\gamma^\uparrow$ and $\gamma^\downarrow$ are $4\times 4$ matrices $(\gamma^\uparrow)^{\alpha}_a$, $(\gamma^\downarrow)^a_{\alpha}$. Since they are ghosts for odd generators, they are even variables.

We will use an auxiliary grading preserved by the Chevalley--Eilenberg differential, the ``bonus symmetry'' of \cite{Intriligator:1998ig}. It is defined for any representation $V$ which can be extended to a representation
of $\mathfrak{pgl}(4|4)$. In other words, we can define $\rho(\Sigma) \in \operatorname{End}(V)$ so that
\begin{equation*}
    [\rho(\Sigma),\rho(\xi)]
        = \rho([\Sigma,\xi])
\end{equation*}
for all $\xi\in \mathfrak{g}$. Define
\begin{align*}
    Y &\in \operatorname{End}\left(C^p(\mathfrak{g},V)\right)
    \\
    Y &= (\gamma^\uparrow)^{\alpha}_a \frac{\partial}{\partial (\gamma^\uparrow)^{\alpha}_a}
    - (\gamma^\downarrow)^a_{\alpha} \frac{\partial}{\partial (\gamma^\downarrow)^a_{\alpha}}
    - \frac{1}{2}[\Sigma,-] ;
\end{align*}
it commutes with $d$:
\begin{equation} \label{RChargeConservation}
    [Y,d] = 0 .
\end{equation}
Indeed, let us consider the complex of $\mathfrak{pgl}(4|4) \supset \mathfrak{psl}(4|4)$. This corresponds to adding additional ghost $c_R$ (corresponding to $\Sigma$). The coefficient of $c_R$ in $d_{\mathfrak{pgl}(4|4)}^2 = 0$ is Eq. (\ref{RChargeConservation}).

\subsection{Hochschild-Serre spectral sequence}\label{sec:HochschildSerre}

Consider the even subalgebra
\begin{align*}
 \mathfrak{g}_{\text{even}}
    = \mathfrak{sl}(4) \oplus \mathfrak{sl}(4)
    \subset \mathfrak{g} .
\end{align*}
For a $\mathfrak{g}$-module $V$, the Hochschild--Serre spectral sequence associated to $\mathfrak{g}_{\text{even}} \subset \mathfrak{g}$ computes $H^{\bullet}(\mathfrak{g};V)$ --- i.e. the cohomology of $\mathfrak{g}$ with coefficients in $V$. The index $p$ counts the number of odd ghost ($\gamma$) inputs; $q$ is the cohomological degree with respect to $\mathfrak{g}_{\text{even}}$. The first page is given by
\begin{equation*}
    {\cal E}_1^{p,q}(V) = H^q\bigl( \mathfrak{sl}(4)\oplus\mathfrak{sl}(4) ; \mathrm{S}^p (\mathfrak{g}_{\text{odd}}{}^*) \otimes V \bigr) .
\end{equation*}

\subsection{Cohomology in the trivial representation}\label{sec:CohomologyInTrivialRep}

\subsubsection{Zeroth and first cohomologies}
\label{sec:H0andH1}

For the trivial $\mathfrak{g}$-module $V=\mathbb{C}$, the degree-one terms are absent,
\begin{equation}\label{HSE11}
    {\cal E}_1^{1,0}(\mathbb{C})
        = {\cal E}_1^{0,1}(\mathbb{C})
        = 0 .
\end{equation}
Indeed, there is no $\mathfrak{sl}(4) \oplus \mathfrak{sl}(4)$-invariant linear function on $\mathfrak{g}_{\text{odd}}$, and $H^1\bigl( \mathfrak{sl}(4)\oplus\mathfrak{sl}(4);\mathbb C \bigr)=0$. Hence
\begin{align*}
    H^0_\mathfrak{g}(\mathbb{C}) &= \mathbb{C} , \\
    H^1_\mathfrak{g}(\mathbb{C}) &= 0 .
\end{align*}

% \subsubsection{$H^2(\mathfrak{g};\mathbb{C})$} 
\subsubsection{Second cohomology} \label{sec:SecondCohomologyG}

The only contribution in total degree two is the invariant quadratic polynomial in the odd ghosts
\begin{align*}
    \mathcal{E}_1^{0,2} (\mathbb{C})
        &= {\cal E}_1^{1,1} (\mathbb{C}) = 0 ,
    \\
    \mathcal{E}_1^{2,0}(\mathbb{C})
        &= \mathbb{C}
        \ : \ \gamma^{\alpha}_a \gamma^a_{\alpha} .
\end{align*}
The vanishing in the first line follows from the semisimplicity of $\mathfrak{sl}(4) \oplus \mathfrak{sl}(4)$
(by Whitehead's lemmas) and the absence of invariant linear odd polynomials. The quadratic invariant is the contraction between $\mathfrak{g}_\uparrow$ and $\mathfrak{g}_\downarrow$; it has bonus charge zero. It survives to ${\cal E}_{\infty}^{2,0}$: an incoming $d_1$ would require a nonzero ${\cal E}_1^{1,0}=0$, an incoming $d_2$ would require a nonzero ${\cal E}_1^{0,1} = 0$, and all higher incoming differentials would require negative $p$.

This is the central extension
\begin{align*}
    c_2 &\in H^2_\mathfrak{g} (\mathbb{C}) \\
    c_2 &= \operatorname{Str}(\Sigma \gamma^2) = 2\operatorname{Tr}(\gamma_\uparrow\gamma_\downarrow)
\end{align*}
corresponding to the obvious short exact sequence
\begin{equation*}
    0 \longrightarrow
        \mathbb{C} \longrightarrow
        \mathfrak{sl}(4|4) \longrightarrow
        \mathfrak{psl}(4|4) \longrightarrow
        0 .
\end{equation*}

%\subsubsection{$H^3(\mathfrak{g};\mathbb{C})$} 
\subsubsection{Third cohomology} \label{sec:ThirdCohomologyG}

Among the groups ${\cal E}_1^{p,q}$ with $p+q=3$, the only nonzero one is
\begin{equation*}
    \mathcal{E}_1^{0,3}(\mathbb{C})
        = H^3\bigl( \mathfrak{sl}(4) \oplus \mathfrak{sl}(4) \bigr)
        \cong \mathbb{C}^2 .
\end{equation*}
The two generators are the standard degree-three primitive classes of the two $\mathfrak{sl}(4)$ factors;
both have $R=0$.

The first possible differential which can act on ${\cal E}_4^{0,3}$ is the transgression
\begin{equation} \label{TransgressionD4}
    d_4 \colon {\cal E}_4^{0,3}(\mathbb{C}) \longrightarrow {\cal E}_4^{4,0}(\mathbb{C}) .
\end{equation}
Since $d_4$ preserves the $\mathfrak{u}(1)_Y$-charge, and the source ${\cal E}_4^{0,3}$ consists entirely of classes with zero $\mathfrak{u}(1)_Y$ charges, the image of the transgression is a subspace of the $Y=0$ subspace of $\mathcal{E}_4^{4,0}$, which is two-dimensional (see Eq. \eqref{QuarticInvariants} below). A direct computation shows that $d_4$ has rank one on this two-dimensional space,
so its kernel in ${\cal E}_4^{0,3}$ is one-dimensional. The surviving linear combination is
\begin{equation*}
    c_3 \equiv \operatorname{Str}(C^3) ;
\end{equation*}
here $C$ denotes the full Faddeev--Popov ghost of $\mathfrak{g}$. The supertrace picks the combination of the two even cubic cocycles, which is compatible with the odd brackets.

% \subsubsection{$H^4(\mathfrak{g};\mathbb{C})$} 
\subsubsection{Fourth cohomology} \label{sec:FourthCohomologyG}

Among the groups ${\cal E}_1^{p,q}$ with $p+q=4$, the only nonzero one is
\begin{equation*}
    {\cal E}_1^{4,0}(\mathbb{C})
        = \operatorname{Hom}_{\mathfrak{sl}(4)\oplus\mathfrak{sl}(4)}(\mathrm{S}^4\mathfrak{g}_{\text{odd}},\mathbb{C}) 
        \cong \mathbb{C}^4 .
\end{equation*}
It is spanned by four quartic invariants, which decompose according to $\mathfrak{u}(1)_Y$-charge as
\begin{equation} \label{QuarticInvariants}
    \underbrace{
        (\gamma^{\alpha}_a \gamma^a_{\alpha})^2
        , \quad
        \gamma^{\alpha}_a \gamma^a_{\beta} \gamma^{\beta}_b \gamma^b_{\alpha}
        }_{R = 0}
    , \qquad
    \underbrace{
        \epsilon_{\alpha_1\cdots\alpha_4} \epsilon^{a_1\cdots a_4} \gamma^{\alpha_1}_{a_1} \gamma^{\alpha_2}_{a_2} \gamma^{\alpha_3}_{a_3} \gamma^{\alpha_4}_{a_4}
    }_{\det\gamma^\uparrow,\; R=+4}
    , \qquad
    \underbrace{
        \epsilon^{\alpha_1\cdots\alpha_4} \epsilon_{a_1\cdots a_4} \gamma_{\alpha_1}^{a_1} \gamma_{\alpha_2}^{a_2} \gamma_{\alpha_3}^{a_3} \gamma_{\alpha_4}^{a_4}
    }_{\det\gamma^\downarrow,\; R=-4} .
\end{equation}
The first two are the two trace invariants of $\gamma^\uparrow\gamma^\downarrow$; the last two are
$\det\gamma^\uparrow$ and $\det\gamma^\downarrow$.

Since the transgression $d_4 \colon \mathcal{E}_4^{0,3}(\mathbb{C}) \rightarrow \mathcal{E}_4^{4,0}(\mathbb{C})$ maps entirely within the $Y=0$ sector, $\det\gamma^\uparrow$ ($Y=+4$) and $\det\gamma^\downarrow$ ($Y=-4$) survive on $\mathcal{E}_{\infty}$. Within the $Y=0$ sector, the transgression Eq. (\ref{TransgressionD4}) hits one linear combination. With the above choice of basis, the result is:

\begin{itemize}
    \item $(\gamma^{\alpha}_a \gamma^a_{\alpha})^2$ corresponds on ${\cal E}_{\infty}$ to $c_2^2\in H^4_{\mathfrak{g}}(\mathbb{C})$.
    \item $\gamma^{\alpha}_a \gamma^a_{\beta} \gamma^{\beta}_b \gamma^b_{\alpha}$ is in the image of $d_4$ of Eq. (\ref{TransgressionD4}).
    \item $\det\gamma^\uparrow$ and $\det\gamma^\downarrow$ survive to ${\cal E}_\infty^{4,0}$ by $\mathfrak{u}(1)_Y$-charge conservation, giving two further independent classes.
\end{itemize}

Therefore
\begin{equation*}
    H^4_{\mathfrak{g}}(\mathbb{C})
    \ \colon \
    c_2^2, \ \det\gamma^\uparrow, \ \det\gamma^\downarrow .
\end{equation*}

\subsection{Cohomology in the adjoint representation}\label{sec:CohomologyInAdjointRep}

Now consider $V=\mathrm{ad}$, i.e. the adjoint representation of $\mathfrak{g}$ on itself. The first low-degree terms are:
\begin{align*}
    \mathcal{E}_1^{0,1}(\mathrm{ad}) &= 0 ,
    \\
    \mathcal{E}_1^{1,0}(\mathrm{ad}) &= H^0_{\mathfrak{sl}(4) \oplus \mathfrak{sl}(4)}(\operatorname{Hom}(\mathfrak{g}_{\text{odd}},\mathfrak{g})) \cong \mathbb{C}^2 .
\end{align*}
The two invariant linear maps $\mathfrak{g}_{\text{odd}} \longrightarrow \mathfrak{g}$ are generated by the identity
map (i.e. $\gamma$) and by $[\Sigma,\gamma]$ (see Eq. \eqref{DefSigma}). The latter is the odd part of the outer grading derivation. Both have $Y = 0$. Similarly:
\begin{align*}
    \mathcal{E}_1^{0,2}(\mathrm{ad})
        &= \mathcal{E}_1^{1,1}(\mathrm{ad}) = 0 ,
    \\
    \mathcal{E}_1^{2,0}(\mathrm{ad})
        &= \operatorname{Hom}_{\mathfrak{sl}(4) \oplus \mathfrak{sl}(4)}(\mathrm{S}^2 \mathfrak{g}_{\text{odd}} , \mathfrak{g} )
        \cong \mathbb{C}^2 ,
\end{align*}
where $\mathcal{E}_1^{2,0}(\mathrm{ad})$ is generated by $(\gamma^\uparrow\gamma^\downarrow)_0$ and $(\gamma^\downarrow\gamma^\uparrow)_0$, both of $\mathfrak{u}(1)_Y$-charge zero, where $(-)_0$ is the projection of $\mathfrak{gl}(4)$ onto $\mathfrak{sl}(4)$ by removing the trace, i.e.
\begin{equation*}
    (X)_0 \equiv X - \frac{1}{4} \operatorname{tr}(X) \, \mathbb{I}_4 .
\end{equation*}

In degree three, we have
\begin{align*}
    \mathcal{E}_1^{0,3}(\mathrm{ad})
        &= \mathcal{E}_1^{1,2}(\mathrm{ad})
        = \mathcal{E}_1^{2,1}(\mathrm{ad}) = 0 ,
    \\
    \mathcal{E}_1^{3,0}(\mathrm{ad})
        &= \operatorname{Hom}_{\mathfrak{sl}(4) \oplus \mathfrak{sl}(4)} (\mathrm{S}^3 \mathfrak{g}_{\text{odd}} , \mathfrak{g})
        \cong \mathbb{C}^6 ;
\end{align*}
the six independent invariant cubic maps are:
\begin{align*}
    & \gamma^\uparrow \gamma^\downarrow \gamma^\uparrow ,
    \\
    & \gamma^\downarrow \gamma^\uparrow \gamma^\downarrow ,
    \\
    & \operatorname{Tr}(\gamma^\uparrow \gamma^\downarrow) \, \gamma^\uparrow ,
    \\
    & \operatorname{Tr}(\gamma^\uparrow \gamma^\downarrow) \, \gamma^\downarrow ,
    \\
    \mu_3^\uparrow \equiv\;
    & \epsilon^{\alpha_1 \alpha_2 \alpha_3 \alpha_4} \epsilon_{a_1 a_2 a_3 a_4} \gamma^\uparrow{}^{a_1}_{\alpha_1} \gamma^\uparrow{}^{a_2}_{\alpha_2} \gamma^\uparrow{}^{a_3}_{\alpha_3} E^{a_4}_{\alpha_4} ,
    \\
    \mu_3^\downarrow \equiv\;
    & \epsilon_{\alpha_1 \alpha_2 \alpha_3 \alpha_4} \epsilon^{a_1 a_2 a_3 a_4} \gamma^\downarrow{}_{a_1}^{\alpha_1} \gamma^\downarrow{}_{a_2}^{\alpha_2} \gamma^\downarrow{}_{a_3}^{\alpha_3} E_{a_4}^{\alpha_4} ,
\end{align*}
where $E^a_{\alpha}$ is the $4\times 4$ matrix filled with zeroes except on the entry $(a,\alpha)$, where it is $1$ --- in other words, $\{ E^a_\alpha \}_{a,\alpha}$ is the canonical basis of $\operatorname{Hom}(\mathbb{C}^4,\mathbb{C}^4) \cong \mathfrak{gl}(4)$.

The last two maps, $\mu_3^\uparrow$ and $\mu_3^\downarrow$, are the adjugate matrices of $\gamma^\uparrow$ and $\gamma^\downarrow$, respectively. As a mnemonic, one can interpret them as
\begin{align*}
    \mu_3^\uparrow
        &= 6\,(\det\gamma^\uparrow) \, (\gamma^\uparrow)^{-1} 
        \quad\text{when}\ \det\gamma^\uparrow\neq 0 ,
    \\
    \mu_3^\downarrow
        &= 6\,(\det\gamma^\downarrow) \, (\gamma^\downarrow)^{-1} 
        \quad\text{when}\ \det\gamma^\downarrow\neq 0 ;
\end{align*}
recall, however, that the polynomial definition (i.e. the one of the form $\epsilon\epsilon\gamma\gamma\gamma E$) above is valid even when the determinant vanishes.

These six generators split by $\mathfrak{u}(1)_Y$-charge in the following way:
\begin{align*}
    Y = 0 \ : & \quad
    \gamma^\uparrow \gamma^\downarrow \gamma^\uparrow
    , \ 
    \gamma^\downarrow \gamma^\uparrow \gamma^\downarrow
    , \ 
    \operatorname{Tr}(\gamma^\uparrow \gamma^\downarrow) \, \gamma^\uparrow
    , \
    \operatorname{Tr}(\gamma^\uparrow \gamma^\downarrow) \, \gamma^\downarrow ,
    \\
    Y = +2 \ : & \quad \mu_3^\uparrow ,
    \\
    Y = -2 \ : & \quad \mu_3^\downarrow .
\end{align*}
Since all degree-two generators have $Y=0$, the adjugates $\mu_3^{\uparrow,\downarrow}$ are not $d_1$-exact. For $p+q < 4$, the spectral sequence stabilizes at the second page, and the only nonzero entries are in the row $q=0$. Thus $\mathcal{E}_2^{p,0}$ is the relative cohomology $H^p(\mathfrak{g},\mathfrak{g}_{\rm even};\mathrm{ad})$, i.e. the cohomology of the complex
\begin{equation*}
    0
    \longrightarrow
        \underbrace{{\cal E}_1^{0,0}(\mathrm{ad})}_{0}
    \stackrel{d_1}{\longrightarrow}
        \underbrace{{\cal E}_1^{1,0}(\mathrm{ad})}_{\mathbb{C}^2}
    \stackrel{d_1}{\longrightarrow}
        \underbrace{{\cal E}_1^{2,0}(\mathrm{ad})}_{\mathbb{C}^2}
    \stackrel{d_1}{\longrightarrow}
        \underbrace{{\cal E}_1^{3,0}(\mathrm{ad})}_{\mathbb{C}^6}
    \stackrel{d_1}{\longrightarrow}
    \cdots .
\end{equation*}
Evaluating $d_1$ on the invariant generators gives the following cohomology. In degree one, $\gamma$ is not closed, while the grading derivation $[\Sigma,\gamma]$ is closed. In degree three, the two adjugate-type maps $\mu_3^\uparrow$ and $\mu_3^\downarrow$ survive by the conservation of $\mathfrak{u}(1)_Y$-charge as argued above, and the explicit $d_1$ computation shows that, among the four $Y=0$
generators, the combination $c_2 \, \mu_1$ also survives, i.e.
\begin{align*}
    \mathcal{E}_2^{0,0}(\mathrm{ad}) &= 0 ,
    \\
    \mathcal{E}_2^{1,0}(\mathrm{ad}) &\cong \mathbb{C}
    \ \colon \ \mu_1 = [\Sigma,\gamma] ,
    \\
    \mathcal{E}_2^{2,0}(\mathrm{ad}) &= 0 ,
    \\
    \mathcal{E}_2^{3,0}(\mathrm{ad}) &\cong \mathbb{C}^3
    \ \colon \ c_2\mu_1 , \ \mu_3^\uparrow , \ \mu_3^\downarrow .
\end{align*}

\section{Second cohomology in ghost number two vertices} \label{SecondCohomology}

Here we will use the spectral sequence of $\mathfrak{g}_{\text{even}} \subset \mathfrak{g}$ to study $H^n(\mathfrak{g};H^2_Q)$ --- i.e. the $n$-th cohomology of $\mathfrak{g}$ with coefficients in $H^2_Q$ (the second cohomology of $Q$).

Consider ${\cal E}_1^{0,1}(H^2_Q) = H^1(\mathfrak{g}_{\text{even}};H^2_Q)$. We \emph{conjecture} that $H^1(\mathfrak{g}_{\text{even}};H^2_Q) = 0$; however, we do not have a rigorous computation of this cohomology. We argue that it vanishes already in flat space. Our intuition is that the only nontrivial ghost number one cohomology class of $\mathfrak{g}$ with coefficients in linearized SUGRA solutions in flat space corresponds to linear axion and dilaton solutions of the form
\begin{align*}
    \phi &= a \; c_{\mu} x^{\mu} ,
    \\
    C &= b \; c_{\mu} x^{\mu} ,
\end{align*}
for constant $a$ and $b$. They correspond to the image under $d_{\text{Lie}}$ of the field configurations satisfying $\square \phi = \text{const}$. This however is not annihilated by $d_1 \colon {\cal E}_1^{0,1}(H^2_Q) \longrightarrow {\cal E}_1^{1,1}(H^2_Q)$.

In the case of flat space, the image of $d_1$ is the dilatino profile
\begin{equation*}
    \boldsymbol{\lambda}
        = (a + \mathrm{i}b) c^{\mu} \, \Gamma_{\mu}\gamma ;
\end{equation*}
we follow \cite{Green:1987mn} in denoting $\boldsymbol{\lambda}$ the dilatino field of the Type IIB SUGRA; it is a complex Weyl spinor. Remember that we consider all $H_Q^2$, including the non-physical vertices. Therefore, there is also a class corresponding to the
nonphysical linear dilaton profile, which takes the form
\begin{align*}
    \phi_{\text{non-phys}}
    &= a' \; c_{\mu} x^{\mu} .
\end{align*}
Its image under $d_1$ is the non-physical dilatino profile $a' \, c^{\mu}\Gamma_{\mu}\gamma$. Thus
\begin{equation} \label{E210Vanishing}
    {\cal E}_2^{0,1}(H^2_Q) = {\cal E}_2^{1,1}(H^2_Q)
    = 0 .
\end{equation}

We conjecture:
\begin{align} \label{E102Vanishing}
    {\cal E}_1^{0,2}(H^2_Q) = 0 .
\end{align}
The bottom row of the second page is the relative cohomology
\begin{equation*}
    {\cal E}_2^{p,0}(H^2_Q)
        = H^p(\mathfrak{g},\mathfrak{g}_{\rm even}; H^2_Q) .
\end{equation*}
The relative cohomology complex is built on $\mathfrak{g}_{\text{even}}$-invariant cochains. In particular, all cochains take values in finite-dimensional subspaces of $H^2_Q$.

Consider the following two classes:
\begin{align}
    & (\Lambda \circ \mu_1)^2 , \label{LambdaCircMu1}
    \\
    & c_2 \, \operatorname{Str}(\lambda_{\text{L}} \lambda_{\text{R}}) . \label{C2Dil0}
\end{align}
We conjecture that these two cohomology classes generate
$H^2(\mathfrak{g},\mathfrak{g}_{\text{even}}; H^2_Q)$; thus, by Eqs. (\ref{E210Vanishing}) and (\ref{E102Vanishing}), they generate the whole space $H^2(\mathfrak{g};H^2_Q)$. The class $(\Lambda \circ \mu_1)^2$ takes values (``coefficients'') in the space of linearized beta-deformations.

There is an exact sequence of $\mathfrak{g}$-modules, including the physical beta-deformation $\beta$, and $\beta'$ without zero trace conditions (see Section \ref{sec:DilatonAndBeta}), namely
\begin{equation*}
    0   \longrightarrow \beta
        \longrightarrow \beta'
        \stackrel{[-,-]}{\longrightarrow}
        \mathrm{ad} \longrightarrow 0 ,
\end{equation*}
and the corresponding exact sequence
\begin{equation*}
    \cdots
        \longrightarrow H_{\mathfrak{g}}^1(\beta')
        \longrightarrow \underbrace{H_{\mathfrak{g}}^1(\mathrm{ad})}_{\mathbb{C}}
        \longrightarrow H_{\mathfrak{g}}^2(\beta)
        \longrightarrow H_{\mathfrak{g}}^2(\beta')
        \longrightarrow \underbrace{H^2_{\mathfrak{g}}(\mathrm{ad})}_{0}
        \longrightarrow \cdots .
\end{equation*}
For the physical $\beta$,
\begin{equation*}
    H^2_{\mathfrak{g}}(\beta) \cong \mathbb{C}^2 ;
\end{equation*}
one class is
\begin{equation*}
    [\Sigma,C] \wedge [\Sigma,C] - d_{\text{Lie}}[-,-]^{-1} \bigl[\Sigma,[\Sigma,C]\bigr]
    \quad\in H^2_{\mathfrak{g}}(\beta) .
\end{equation*}
It is the sum of a class of terms of the type $t^{[A}_{(C}\otimes t^{B]}_{D)}$ and terms of the type $t^{(A}_{[C}\otimes t^{B)}_{D]}$. Changing the relative sign of the two types of terms, we obtain the second class in $H^2_{\mathfrak{g}}(\beta)$. Here, since we consider all $H_Q^2$ including the non-physical vertices, there is only one
cohomology class with the coefficients in the beta-deformation multiplet, namely $(\Lambda\circ \mu_1)^2$.

\section{Second page for \texorpdfstring{$\boldsymbol{\mathcal{R} ={\mathbb{C}}}$}{cal(R) = ℂ}} \label{Grids}

The tables below show grids for $E_2$ in the trivial representation. Tables \ref{tab:E2} and \ref{tab:E2tilde} show the spaces on the corner of the $E_2$ and $\widetilde{E}_2$, respectively. Tables \ref{tab:E2gen} and \ref{tab:E2tildegen} show the generators of those spaces; spaces in gray are not computed in this paper. We first show diagrams, unfolding the computation in three steps, and then explain the details.

\begin{table}[!ht]
\centering
\renewcommand{\arraystretch}{1.5}
\begin{tabular}{|c||*{5}{>{\centering\arraybackslash}m{2.5cm}|}}
    \hline
    $\boldsymbol{4}$
        & $0$ & $0$ & $0$ & $0$ & $0$
    \\ \hline
    $\boldsymbol{3}$
        & $H^0(\mathfrak{g};H_Q^3)$
         & $H^1(\mathfrak{g};H_Q^3)$ 
          & $H^2(\mathfrak{g};H_Q^3)$ 
           & $H^3(\mathfrak{g};H_Q^3)$
            & $H^4(\mathfrak{g};H_Q^3)$
    \\ \hline
    $\boldsymbol{2}$
        & $H^0(\mathfrak{g};H_Q^2)$
         & $H^1(\mathfrak{g};H_Q^2)$
          & $H^2(\mathfrak{g};H_Q^2)$
           & $H^3(\mathfrak{g};H_Q^2)$
            & $H^4(\mathfrak{g};H_Q^2)$
    \\ \hline
    $\boldsymbol{1}$
        & $0$
         & $H^1(\mathfrak{g};\mathrm{ad})$
          & $0$
           & $H^3(\mathfrak{g};\mathrm{ad})$
            & $H^4(\mathfrak{g};\mathrm{ad})$
    \\ \hline
    $\boldsymbol{0}$
        & $\mathbb{C}$
         & $0$
          & $\mathbb{C}$
           & $H^3(\mathfrak{g};\mathbb{C})$
            & $H^4(\mathfrak{g};\mathbb{C})$
    \\ \hline\hline
    \diagbox[
        dir=SW,
        innerwidth=0.75cm, 
        innerleftsep = 0.125cm,
        innerrightsep = 0.125cm,
        height=1.75\line,
        ]{$\boldsymbol{q}$}{$\boldsymbol{p}$}
        & $\boldsymbol{0}$
         & $\boldsymbol{1}$ 
          & $\boldsymbol{2}$ 
           & $\boldsymbol{3}$ 
            & $\boldsymbol{4}$
    \\ \hline
\end{tabular}
\caption{The spaces in the corner of the second page $E_2^{p,q} = H^p_\mathfrak{g} \bigl( H^q_Q(\operatorname{Coind}^G_H \mathcal{P}^\bullet) \bigr)$.}
\label{tab:E2}
\end{table}
\begin{table}[ht]
\centering
\renewcommand{\arraystretch}{1.5}
\begin{tabular}{|c||*{5}{>{\centering\arraybackslash}m{2.5cm}|}}
    \hline
    $\boldsymbol{4}$
        & $0$ & $0$ & $0$ & $0$ & $0$
    \\ \hline
    $\boldsymbol{3}$
        & $H^3(\mathfrak{h}) \otimes (\mathcal{P}^0)^\mathfrak{h}$
         & $0$ 
          & $H^3(\mathfrak{h})\otimes (\mathcal{P}^2)^\mathfrak{h}$ 
           & $0$
            & $H^3(\mathfrak{h}) \otimes (\mathcal{P}^4)^\mathfrak{h}$
    \\ \hline
    $\boldsymbol{2}$
        & $0$ & $0$ & $0$ & $0$ & $0$
    \\ \hline
    $\boldsymbol{1}$
        & $0$ & $0$ & $0$ & $0$ & $0$
    \\ \hline
    $\boldsymbol{0}$
        & $(\mathcal{P}^0)^\mathfrak{h}$
         & $0$
          & $(\mathcal{P}^2)^\mathfrak{h}$
           & $0$
            & $(\mathcal{P}^4)^\mathfrak{h}$
    \\ \hline\hline
    \diagbox[
        dir=SW,
        innerwidth=0.75cm, 
        innerleftsep = 0.125cm,
        innerrightsep = 0.125cm,
        height=1.75\line,
        ]{$\boldsymbol{q}$}{$\boldsymbol{p}$}
        & $\boldsymbol{0}$
         & $\boldsymbol{1}$ 
          & $\boldsymbol{2}$ 
           & $\boldsymbol{3}$ 
            & $\boldsymbol{4}$
    \\ \hline
\end{tabular}
\caption{The spaces in the corner of the second page $\widetilde{E}_2^{p,q} = H^p_Q \bigl( H^q_\mathfrak{g} ( \operatorname{Coind}^{\mathfrak{g}}_{\mathfrak{h}} \mathcal{P}^\bullet) \bigr)$.}
\label{tab:E2tilde}
\end{table}
\begin{table}[ht]
\centering
\renewcommand{\arraystretch}{1.5}
\begin{tabular}{|c||*{5}{>{\centering\arraybackslash}m{2.5cm}|}}
    \hline
    $\boldsymbol{4}$
        & $0$ & $0$ & $0$ & $0$ & $0$
    \\ \hline
    $\boldsymbol{3}$
        & $\bullet\, \scalebox{.75}{$W_{\text{dilaton} \atop \text{0-mode}}$}$,
            $\bullet\, \scalebox{.75}{$W_{\text{axion} \atop \text{0-mode}}$}$
         & $0$ 
          & ${\color{gray} H^2(\mathfrak{g};H_Q^3)}$ 
           & ${\color{gray} H^3(\mathfrak{g};H_Q^3)}$
            & ${\color{gray} H^4(\mathfrak{g};H_Q^3)}$
    \\ \hline
    $\boldsymbol{2}$
        & $\bullet\, \operatorname{STr}(\lambda_{\text{L}} \lambda_{\text{R}})$
         & $0$
          & $\bullet\, \scalebox{.75}{$(\Lambda\circ\mu_1)^2$}$,
            $\bullet\, \scalebox{.75}{$c_2 \, \operatorname{Str}(\lambda_{\text{L}} \lambda_{\text{R}})$}$
           & ${\color{gray} H^3(\mathfrak{g};H_Q^2)}$
            & ${\color{gray} H^4(\mathfrak{g};H_Q^2)}$
    \\ \hline
    $\boldsymbol{1}$
        & $0$
         & $\bullet\, \Lambda\circ\mu_1$
          & $0$
           & $\bullet\, \scalebox{0.75}{$c_2 \, (\Lambda\circ\mu_1)$}$,
                $\bullet\, \scalebox{0.75}{$\Lambda\circ\mu_3^\uparrow$}$,
                $\bullet\, \scalebox{0.75}{$\Lambda \circ \mu_3^\downarrow$}$
            & ${\color{gray}H^4(\mathfrak{g};\mathrm{ad})}$
    \\ \hline
    $\boldsymbol{0}$
        & $\mathbb{C}$
         & $0$
          & $\bullet\, c_2$
           & $\bullet\, c_3$
            & $\bullet\, \scalebox{.75}{$\det\gamma^\uparrow$}$,
                $\bullet\, \scalebox{.75}{$\det\gamma^\downarrow$}$,
                $\bullet\, \scalebox{0.75}{$(c_2)^2$}$
    \\ \hline\hline
    \diagbox[
        dir=SW,
        innerwidth=0.75cm, 
        innerleftsep = 0.125cm,
        innerrightsep = 0.125cm,
        height=1.75\line,
        ]{$\boldsymbol{q}$}{$\boldsymbol{p}$}
        & $\boldsymbol{0}$
         & $\boldsymbol{1}$ 
          & $\boldsymbol{2}$ 
           & $\boldsymbol{3}$ 
            & $\boldsymbol{4}$
    \\ \hline
\end{tabular}
\caption{Generators in the corner of the second page $E_2^{p,q} = H^p_\mathfrak{g} \bigl( H^q_Q (\operatorname{Coind}^G_H \mathcal{P}^\bullet) \bigr)$.}
\label{tab:E2gen}
\end{table}
\begin{table}[!ht]
\centering
\renewcommand{\arraystretch}{1.5}
\begin{tabular}{|c||*{5}{>{\centering\arraybackslash}m{2.5cm}|}}
    \hline
    $\boldsymbol{4}$
        & $0$ & $0$ & $0$ & $0$ & $0$
    \\ \hline
    $\boldsymbol{3}$
        & $\bullet\, \scalebox{.75}{$\operatorname{STr}(\mathrm{e}^{-\omega} d\mathrm{e}^{\omega})^3$}$, 
            $\bullet\, \scalebox{.75}{$\operatorname{Tr}(\mathrm{e}^{-\omega} d\mathrm{e}^\omega)^3$}$
         & $0$ 
          & \scalebox{.85}{$\color{gray} H^3(\mathfrak{h})\otimes (\mathcal{P}^2)^\mathfrak{h}$} 
           & $0$
            & \scalebox{.85}{$\color{gray} H^3(\mathfrak{h}) \otimes (\mathcal{P}^4)^\mathfrak{h}$}
    \\ \hline
    $\boldsymbol{2}$
        & $0$ & $0$ & $0$ & $0$ & $0$
    \\ \hline
    $\boldsymbol{1}$
        & $0$ & $0$ & $0$ & $0$ & $0$
    \\ \hline
    $\boldsymbol{0}$
        & $\mathbb{C}$
         & $0$
          & $\bullet\, \scalebox{.75}{$\operatorname{STr}(\lambda_{\text{L}} \lambda_{\text{R}})$}$,
            $\bullet\, \scalebox{.75}{$\lambda_{\text{L}}^2$}$,
            $\bullet\, \scalebox{.75}{$\lambda_{\text{R}}^2$}$
           & $0$
            & \scalebox{.85}{$\mathbb{C}^7$,}
            \scalebox{.85}{see Eq. \eqref{E2tilde40generators}}
    \\ \hline\hline
    \diagbox[
        dir=SW,
        innerwidth=0.75cm, 
        innerleftsep = 0.125cm,
        innerrightsep = 0.125cm,
        height=1.75\line,
        ]{$\boldsymbol{q}$}{$\boldsymbol{p}$}
        & $\boldsymbol{0}$
         & $\boldsymbol{1}$ 
          & $\boldsymbol{2}$ 
           & $\boldsymbol{3}$ 
            & $\boldsymbol{4}$
    \\ \hline
\end{tabular}
\caption{Generators in the corner of the second page $\widetilde{E}_2^{p,q} \cong H^p_Q \bigl( H^q(\mathfrak{h}) \otimes (\mathcal{P}^\bullet)^\mathfrak{h} \bigr)$.}
\label{tab:E2tildegen}
\end{table}

\subsection{Degree two}\label{DegreeTwo}

At the degree two, the resolution of BRST-exact expressions $\lambda_{\text{L}}^2$ and $\lambda_{\text{R}}^2$ has been worked out, essentially in \cite{Berkovits:2011kn}. Of course, they cannot be resolved covariantly, because these expressions are $\mathfrak{g}$-invariant. Here we will compute the cohomological obstacle to covariant resolution.

We will use the abbreviated notation introduced in (\ref{AbbreviateC2L}) and (\ref{AbbreviateC2R}).

\subsubsection[Structure of \texorpdfstring{$\widetilde{E}_2^{p,2-p}$}{Ẽ2(p,2-p)}]{Structure of \texorpdfstring{$\boldsymbol{\widetilde{E}_2^{p,2-p}}$}{Ẽ2(p,2-p)}} \label{E2Til2}

The only nonzero component is $\widetilde{E}_2^{2,0}$, which is generated by
\begin{equation*}
    \widetilde{E}_2^{2,0} = \mathbb{C}\bigl\langle \lambda_{\text{L}}^2, \lambda_{\text{R}}^2, \operatorname{STr}(\lambda_{\text{L}}\lambda_{\text{R}}) \bigr\rangle .
\end{equation*}

\subsubsection[Structure of \texorpdfstring{$E_2^{p,2-p}$}{E2(p,2-p)}]{Structure of \texorpdfstring{$\boldsymbol{E_2^{p,2-p}}$}{E2(p,2-p)}} \label{E22}

\begin{align}
    E_2^{0,2}
        &= H^0 (\mathfrak{g};H^2_Q)
        \ \colon \ \operatorname{Str}(\lambda_{\text{L}}\lambda_{\text{R}}) , \label{E202}
    \\
    E_2^{1,1}
        &= H^1 (\mathfrak{g};\mathrm{ad})
        \ \colon \ \Lambda\bigl\langle [\Sigma,c] \bigr\rangle , \label{E211}
    \\
    E_2^{2,0}
        &= H^2 (\mathfrak{g},\mathbb{C})
        \ \colon \ c_2 . \label{E220}
\end{align}

\subsubsection[The relation between \texorpdfstring{$\widetilde{E}_2^{p,2-p}$ and $E_2^{p,2-p}$}{Ẽ2(p,2-p) and E2(p,2-p)}]{The relation between \texorpdfstring{$\boldsymbol{\widetilde{E}_2^{p,2-p}}$ and $\boldsymbol{E_2^{p,2-p}}$}{Ẽ2(p,2-p) and E2(p,2-p)}} \label{RelationEAndETil}

\paragraph{The difference} \label{sec:LeftMinusRight}

The difference $\lambda_{\text{L}}^2 - \lambda_{\text{R}}^2$ is related to $H^1_\mathfrak{g}(\mathrm{ad})$ in the following
way:
\begin{align}
    \lambda_{\text{L}}^2 - \lambda_{\text{R}}^2
        &= \frac{1}{8} \,Q \,\operatorname{Str}\left(\Sigma\, g^{-1}(\lambda_{\text{L}} - \lambda_{\text{R}})\right) , \label{ResolutionL2Minus}
    \\
    d_{\mathfrak{g}} \, \operatorname{Str} \left( \Sigma \, g^{-1}(\lambda_{\text{L}} - \lambda_{\text{R}}) \right)
        &= \operatorname{Str}\left( [\Sigma,c] \, g^{-1}(\lambda_{\text{L}} - \lambda_{\text{R}})g \right)
        = \Lambda\bigl\langle [\Sigma,c] \bigr\rangle ;
\end{align}
or yet
\begin{equation*}
    \lambda_{\text{L}}^2 - \lambda_{\text{R}}^2 + \frac{1}{8}\, \Lambda\bigl\langle [\Sigma,c] \bigr\rangle
    =
    (Q + d_{\mathfrak{g}}) \, \frac{1}{8} \, \operatorname{Str} \left( \Sigma \, g^{-1}(\lambda_{\text{L}} - \lambda_{\text{R}}) \right) .
\end{equation*}
This means that both $\lambda_{\text{L}}^2 - \lambda_{\text{R}}^2$ in $\widetilde{E}_2$ and
$\frac{1}{8} \, \Lambda\left\langle[\Sigma,c]\right\rangle\in H^1_{\mathfrak{g}}(\mathrm{ad}) = E_2^{1,1}$ in $E_2$ correspond to the same cohomology class of $d_{\mathfrak{g}} + Q$; hence $\lambda_{\text{L}}^2 - \lambda_{\text{R}}^2$ corresponds to the generator of $H^1_\mathfrak{g}(\mathrm{ad})$.

\paragraph{The sum}\label{sec:LeftPlusRight}

The sum $\lambda_{\text{L}}^2 + \lambda_{\text{R}}^2$ is also BRST-exact. Its resolution corresponds to $H^2(\mathfrak{g};\mathbb{C})$, as we will now explain.

Let ${\cal C}A$ be the pure spinor cone bundle over super-AdS, i.e.
\begin{equation*}
    {\cal C}A = \frac{\widehat{G}\times \text{Cone}_{\text{L}} \times \text{Cone}_{\text{R}}}{H \times \mathrm{U}(1)} .
\end{equation*}
Consider the $\mathrm{U}(1)$ bundle
\begin{gather*}
    \frac{\widehat{G} \times \text{Cone}_{\text{L}} \times \text{Cone}_{\text{R}}}{H}
    \equiv \widehat{{\cal C}A} \longrightarrow {\cal C}A ,
\end{gather*}
and the following $\mathrm{U}(1)$-connection 1-form on that bundle:
\begin{equation} \label{ConnectionForm}
    {\cal A} = \frac{1}{8} \, \operatorname{Str}\left(\Sigma\, g^{-1}dg\right) .
\end{equation}
Notice that $\cal A$ is not $G$-invariant; it is only invariant under the bosonic isometries, but not under the supersymmetries. Its curvature ${\cal F} = d{\cal A}$ is a base form, inducing a two-form on the base ${\cal C}A$:
\begin{align*}
    {\cal F} &\in \Omega^2({\cal C}A)
    \\
    {\cal F} &= -\frac{1}{8} \, \operatorname{Str} \left(\Sigma\, (g^{-1}dg)^2 \right) .
\end{align*}

To better organize the computations, it is sometimes useful to think of the total differential $Q + d_{\mathfrak{g}}$
as the Lie superalgebra differential of the direct sum
$\mathbb{C}^{0|1} \oplus \mathfrak{g}$, where $\mathbb{C}^{0|1}$ is generated by the BRST vector field $Q$ \cite{Bernardes:2023nit}. Let us denote\footnote{Recall that the Faddeev--Popov ghosts of $\mathfrak{g}$ are denoted $C^A$.} $u$ the Faddeev--Popov ghost for $\mathbb{C}^{0|1}$. Being a closed 2-form, $\cal F$ defines a 2-cocycle $\psi$ of $\mathbb{C}^{0|1} \oplus \mathfrak{g}$ in the following way:
\begin{align}
    \psi &\in Z^2\left(\mathbb{C}^{0|1}\oplus \mathfrak{g} ; C^{\infty}({\cal C}A) \right)
    \nonumber \\
    \psi &= \left.\left( \mathrm{e}^{\iota\langle u Q + C^At_A\rangle} {\cal F}\right)\right|_{\text{0-form}}
    \nonumber \\
    &= - u^2(\lambda_{\text{L}}^2 + \lambda_{\text{R}}^2)
        - \frac{1}{8} \, u \, \operatorname{Str}\left(\Sigma \left[\,g^{-1}(\lambda_{\text{L}} + \lambda_{\text{R}}) \, g \,,\, C^A t_A \right]\right) 
        - c_{AB} C^A C^B , \label{SumAndH2}
\end{align}
where $\iota$ is the contraction map of a vector with a 2-form, i.e. $\iota\langle v \rangle \omega = \omega(v,-)$. Here $c_{AB} C^A C^B \in Z^2(\mathfrak{g})$ is the 2-cocycle corresponding to the central extension of $\mathfrak{g}$.

Let us consider a local $\mathrm{U}(1)$ gauge fixing, i.e. an embedding of ${\cal C}A$ into $\widehat{{\cal C}A}$ as a ``gauge slice'',
\begin{equation*}
    {\cal C}A \longrightarrow \widehat{{\cal C}A} .
\end{equation*}
The action of $\mathbb{C}^{0|1} \oplus \mathfrak{g}$ on ${\cal C}A$ defines an action on the gauge slice. In particular $Q$ and $d_{\mathfrak{g}}$ act as
\begin{align*}
    Q \, \widehat{g}
        &= (\lambda_{\text{L}} + \lambda_{\text{R}}) \, \widehat{g} + a \langle\lambda\rangle \, \widehat{g} ,
    \\
    d_{\mathfrak{g}} \, \widehat{g}
        &= \widehat{g}(C^A t_A + C^A b_A) ,
\end{align*}
where $a\langle \lambda\rangle \in \mathbb{C}$ and $b\langle C\rangle\in\mathbb{C}$ are the compensators, which are needed in order to keep $\widehat{g}$ on the gauge slice. For the connection form, given by Eq. (\ref{ConnectionForm}), we observe that
\begin{align*}
    \left.\left( \mathrm{e}^{\iota\langle u Q + C^A t_A\rangle} \left({\cal A}|_{\text{gauge slice}} \right) \right)\right|_{\text{0-form}} 
        &= \alpha + \beta \quad\in C^1\left(\mathbb{C}^{0|1}\oplus \mathfrak{g}\,,\,C^{\infty}({\cal C}A)\right) .
    \\
    \alpha &\in u \, a\langle\lambda\rangle ,
    \\
    \beta &= \frac{1}{8} \, C^A \, \operatorname{Str}\bigl(\Sigma \,(t_A + b_A) \bigr) ,
\end{align*}
where $b_A\in\mathbb R$ is the compensator for the infinitesimal right shift $\delta \widehat{g} = \widehat{g} t_A$.

Thus $\psi$ is exact in the Hochschild--Serre complex
$C^{\bullet}\left(\mathbb{C}^{0|1}\oplus \mathfrak{g}\,,\, C^{\infty}({\cal C}A) \right)$:
\begin{equation*}
    \psi = d_{\mathbb{C}^{0|1} \oplus \mathfrak{g}} (\alpha + \beta) .
\end{equation*}
Then, Eq. (\ref{SumAndH2}) implies that $\lambda_{\text{L}}^2 + \lambda_{\text{R}}^2$ represents the same cohomology class of $d_{\mathfrak{g}} + Q$
as $c_{AB}C^A \, C^B\in H^2_{\mathfrak{g}}(\mathbb{C})$.
This class is visible as $\lambda_{\text{L}}^2 + \lambda_{\text{R}}^2$ in $\widetilde{E}^{2,0}$
and as $c_2$ in $E^{2,0}$.

\subsection{Degree three}\label{DegreeThree}

\subsubsection[Structure of \texorpdfstring{$\widetilde{E}_2^{p,3-p}$}{Ẽ2(p,3-p)}]{Structure of \texorpdfstring{$\boldsymbol{\widetilde{E}_2^{p,3-p}}$}{Ẽ2(p,3-p)}} \label{E2Til3}

The only nonzero component is
\begin{equation*}
    \widetilde{E}_2^{0,3} = \mathbb{C} \bigl\langle \operatorname{Str}(\mathrm{e}^{-\omega}de^{\omega})^3, \,\operatorname{Tr}(\mathrm{e}^{-\omega}de^{\omega})^3 \bigr\rangle ,
\end{equation*}
from which we conclude that
\begin{equation}\label{DimETil23}
    \dim \bigoplus_{p=0}^3 \widetilde{E}_2^{p,3-p} = 2 .
\end{equation}

\subsubsection[Structure of \texorpdfstring{$E_2^{p,3-p}$}{E2(p,3-p)}]{Structure of \texorpdfstring{$\boldsymbol{E_2^{p,3-p}}$}{E2(p,3-p)}} \label{E23}

The ghost number three zero-modes generating $E_2^{0,3}$ are
\begin{equation*}
    E_2^{0,3} \cong \mathbb{C}^2
    \ \colon \
    V_{\text{dilaton} \atop \text{0-mode}} ,\; V_{\text{axion} \atop \text{0-mode}} .
\end{equation*}
As $\mathfrak{psu}(2,2|4)$ does not have deformations, we have
\begin{equation*}
    E_2^{2,1} = H^2(\mathfrak{g},\mathrm{ad}) = 0 .
\end{equation*}
In turn, $E_2^{3,0} = H^3(\mathfrak{g})$ is one-dimensional, generated by $\operatorname{STr}\;C^3$:
\begin{equation*}
    E_2^{3,0} \cong \mathbb{C}
    \;\colon\; c_3 .
\end{equation*}

Finally, we conjecture that
\begin{equation}
    E_2^{1,2} = H^1(\mathfrak{g},H_Q^2) = 0 ,
\end{equation}
although we do not have a complete proof of this. From this conjecture, it follows that
\begin{equation}\label{DimE23}
    \dim \bigoplus_{p=0}^3 E_2^{p,3-p} = 3 .
\end{equation}

\subsection{Degree four}\label{DegreeFour}

\subsubsection[Structure of \texorpdfstring{$\widetilde{E}_2^{p,4-p}$}{Ẽ2(p,4-p)}]{Structure of \texorpdfstring{$\boldsymbol{\widetilde{E}_2^{p,4-p}}$}{Ẽ2(p,4-p)}} \label{E2Til4}

The only nonzero square is at $p=4$. It is generated by all $\mathfrak{h}$-invariant polynomials of degree four in $\lambda_{\text{L}}$ and $\lambda_{\text{R}}$. There are seven of them:
\begin{subequations} \label{E2tilde40generators}
\begin{align}
    \widetilde{E}_2^{4,0} \cong \mathbb{C}^7
    \;\colon\;
    & \lambda_{\text{L}}^2 \, \operatorname{Str}(\lambda_{\text{L}}\lambda_{\text{R}})
    ,\; \lambda_{\text{R}}^2 \, \operatorname{Str}(\lambda_{\text{L}}\lambda_{\text{R}})
    , \\
    & (\lambda_{\text{L}}^2)^2
    ,\;(\lambda_{\text{R}}^2)^2
    ,\;\lambda_{\text{L}}^2 \lambda_{\text{R}}^2
    , \\
    & (\operatorname{Str}(\lambda_{\text{L}}\lambda_{\text{R}}))^2
    ,\; \operatorname{Tr}[\lambda_{\text{L}},\lambda_{\text{R}}]^2 .
\end{align}
\end{subequations}
In the language of Section \ref{MatrixRealization}, those are six quadratic polynomials of $\lambda^2$, $\mu^2$ and $\|\lambda\cap\mu\|$, and one more element:
\begin{equation*}
    \big\| \lambda\cap\mu\cup\lambda\cap\mu \bigr\| ,
\end{equation*}
where $\| A^{\cdot\cdot} \| = \omega_{\alpha\beta} A^{\beta\alpha}$ or $\| A_{\cdot\cdot} \| = \omega^{ab} A_{ba}$, depending on the index structure. Therefore
\begin{equation} \label{DimETil24}
    \dim \bigoplus_{p=0}^4 \widetilde{E}_2^{p,4-p} = 7 .
\end{equation}

\subsubsection[Structure of \texorpdfstring{$\widetilde{E}_2^{p,4-p}$}{Ẽ2(p,4-p)}]{Structure of \texorpdfstring{$\boldsymbol{\widetilde{E}_2^{p,4-p}}$}{Ẽ2(p,4-p)}} \label{E24}

The top-left two squares, $E_2^{0,4}$ and $E_2^{1,3}$, are zero
\begin{align}
    E_2^{0,4} &= 0 , \label{E204}
    \\
    E_2^{1,3} &= 0 ; \label{E213}
\end{align}
then follow squares of dimensions 2, 3 and 3,
\begin{align*}
    E_2^{2,2} \cong \mathbb{C}^2 \;\colon\;
        & (\Lambda\circ \mu_1)^2 \\
        & c_2 \operatorname{STr}(\lambda_{\text{L}}\lambda_{\text{R}}) ,
    \\
    E_2^{3,1} \cong \mathbb{C}^3 \;\colon\;
        & c_2 \, \Lambda\circ\mu_1 \\
        & \Lambda \circ \mu_3^\uparrow \\
        & \Lambda \circ \mu_3^\downarrow , \\
    E_2^{4,0} \cong \mathbb{C}^3 \;\colon\;
        & c_2^2 \\
        & \det\gamma^\uparrow \\
        & \det\gamma^\downarrow ;
\end{align*}
therefore
\begin{equation} \label{DimE24}
    \dim \bigoplus_{p=0}^4 E_2^{p,4-p} = 8 .
\end{equation}

\section{Higher differentials for \texorpdfstring{$\boldsymbol{\mathcal{R} = \mathbb{C}}$}{cal(R) = bb(C)}} \label{KnightMove}

To summarize, we have the following dimensions on the second page:
\vspace{8pt}
\begin{center}
    \begin{tabular}{|c||c|c|c|c|c|}
    \hline
    $\boldsymbol{p+q}$
     & $\boldsymbol{0}$
      & $\boldsymbol{1}$
       & $\boldsymbol{2}$
        & $\boldsymbol{3}$ 
         & $\boldsymbol{4}$ \\
    \hline\hline
    $\boldsymbol{\dim E_2}$
     & $1$ & $0$ & $3$ & $3$ & $8$ \\
    \hline
    $\boldsymbol{\dim \widetilde{E}_2}$
     & $1$ & $0$ & $3$ & $2$ & $7$ \\
    \hline
    \end{tabular}
\end{center}

At the level $p+q > 2$, we have not actually computed $d_{\geq 2}$. What we describe here is a {\bf conjecture}. We will argue that, at the infinite page, we have

\begin{center}
    \begin{tabular}{|c||c|c|c|c|c|}
    \hline
    $\boldsymbol{p+q}$
     & $\boldsymbol{0}$
      & $\boldsymbol{1}$
       & $\boldsymbol{2}$
        & $\boldsymbol{3}$ 
         & $\boldsymbol{4}$ \\
    \hline\hline
    $\dim E_{\infty}$            & 1 & 0 & 3 & 1 & 6 \\
    \hline
    $\dim\widetilde{E}_{\infty}$ & 1 & 0 & 3 & 1 & 6 \\
    \hline
\end{tabular}
\end{center}

\subsection[Action of of \texorpdfstring{$d_{\geq 2}$}{d(≥2)}]{Action of of \texorpdfstring{$\boldsymbol{d_{\geq 2}}$}{d(≥2)}} \label{sec:DGeq2}

At the level $p+q=2$, both $\operatorname{Str}(\lambda_{\text{L}} \lambda_{\text{R}})$ and $\Lambda\circ \mu_1$ are
represented by {\bf covariant} vertices, and therefore are annihilated by the higher differentials.

The differential $\tilde{d}_4$ is nonzero, cancelling $\operatorname{Tr}( \mathrm{e}^{-\omega} d\mathrm{e}^{\omega})^3$ against the class in $\widetilde{E}^{4,0}$ represented by $\operatorname{Tr}\{\lambda_{\text{L}},\lambda_{\text{R}}\}^2$. We will now outline the descent procedure for the computation of
$\tilde{d}_4\,\operatorname{Tr}(\mathrm{e}^{-\omega} d\mathrm{e}^{\omega})^3$ using the notations of Section \ref{ShapiroLemma}. We have:
\begin{alignat*}{4}
    Z &= x + \theta ,
    \qquad&\qquad\qquad
    x &\in \mathfrak{g}_{\bar{2}} ,
    \\
    \theta &= \theta_{\text{L}} + \theta_{\text{R}} ,
    \qquad&\qquad\qquad
    \theta_{\text{L}} &\in \mathfrak{g}_{\bar{3}} ,
    \\
    l &= \mathrm{e}^{-\omega} (\lambda_{\text{L}} + \lambda_{\text{R}})\, \mathrm{e}^{\omega} ,
    \qquad&\qquad\qquad
    \theta_{\text{R}} &\in \mathfrak{g}_{\bar{1}} ;
\end{alignat*}
in this language, $d_{\text{Lie}} = d$.

Here are examples of how $Q$ acts:
\begin{align*}
    Q\mathrm{e}^{\omega}
        &= \frac{1}{2} \, \mathrm{e}^{\omega} \, \{l,\theta\}_0 + \ldots ,
    \\
    Q(\mathrm{e}^{-\omega} d\mathrm{e}^{\omega})
        &= - \frac{1}{2} \, \mathrm{e}^{-\omega} \, d\bigl(\mathrm{e}^{\omega} \, \{l,\theta\}_0 \bigr) 
        - \frac{1}{2} \, \{l,\theta\}_0 \, \mathrm{e}^{-\omega} d\mathrm{e}^{\omega} + \ldots
        \\
        &= - \frac{1}{2} \left(d + [e^{-\omega}de^{\omega},-] \right) \{l,\theta\}_0 + \ldots
        \\
        &= - \frac{1}{2} \, \mathrm{e}^{-\omega} d\bigl\{\lambda,\mathrm{e}^{\omega} \theta \mathrm{e}^{-\omega} \bigr\}_0 \mathrm{e}^{\omega} + \ldots ,
\end{align*}
and $\ldots$ are terms of higher order in $x$ and $\theta$.

We will compute the descent of $\operatorname{Tr}(\mathrm{e}^{-\omega} d\mathrm{e}^{\omega})^3$. Let us denote
\begin{align*}
    A &= \bigl( (Q+d) \, \mathrm{e}^{\omega} \bigr) \, \mathrm{e}^{-\omega} ,
    \\
    F &= \bigl( (Q+d)^2 \mathrm{e}^{\omega} \bigr) \, \mathrm{e}^{-\omega}
        = (Q^2 \mathrm{e}^{\omega}) \, \mathrm{e}^{-\omega} 
        = \frac{1}{2} \, \{\lambda_{\text{L}} , \lambda_{\text{R}} \} ;
\end{align*}
we observe that
\begin{equation*}
    (d + Q)A + A^2 = F ,
\end{equation*}
\begin{equation}\label{CSDescent}
    (d + Q)\operatorname{Tr}\left(AF + \frac{1}{3}A^3\right)
        = \operatorname{Tr}\, F^2
        = \operatorname{Tr}\, \{\lambda_{\text{L}} , \lambda_{\text{R}}\}^2 .
\end{equation}
With the grading corresponding to $\widetilde{E}$, the top component of $\operatorname{Tr}\left(AF + \frac{1}{3}A^3\right)$ is $\operatorname{Tr}\left(d\mathrm{e}^{\omega} \mathrm{e}^{-\omega}\right)^3$. Then, Eq. (\ref{CSDescent}) implies that $\operatorname{Tr}\left(d\mathrm{e}^{\omega} \, \mathrm{e}^{-\omega}\right)^3$ cancels with $\operatorname{Tr}\{\lambda_{\text{L}},\lambda_{\text{R}}\}^2$, i.e.
\begin{equation*}
    \widetilde{d}_4 \operatorname{Tr}\left(d\mathrm{e}^{\omega} \mathrm{e}^{-\omega} \right)^3
        = 3\operatorname{Tr}\,\{ \lambda_{\text{L}} , \lambda_{\text{R}} \}^2 .
\end{equation*}
The other class, which is generated by $\operatorname{Str}\left( d\mathrm{e}^{\omega} \right)^3$, survives on $\widetilde{E}^{0,3}$ because
\begin{equation*}
    \operatorname{Str}\, F^2 = 0
\end{equation*}
due to the pure spinor constraints. Indeed, there must be one surviving class in $\widetilde{E}^{p,3-p}$, to match with $c_3 \in E^{3,0}$ --- which survives on $E^{3,0}_{\infty}$. 
(As we have already explained, both
$d_3 \operatorname{Str}(\lambda_{\text{L}}\lambda_{\text{R}})$ and $d_2(\Lambda\circ \mu_1)$ are zero.)

This implies that neither $W_{\text{dilaton} \atop \text{0-mode}}$ nor $W_{\text{axion} \atop \text{0-mode}}$ survive on $E_{\infty}$. In other words, some higher differentials act nonzero on them, cancelling them with something else in
$E^{2,2} \oplus E^{3,1} \oplus E^{4,0}$.

Some information can be gained from considering worldsheet parity, as in Section \ref{FlatSpace}. Notice that $W_{\text{dilaton} \atop \text{0-mode}}$ is parity-even, and its flat space limit is given by Eq. (\ref{GhostNumber3Even}). At the same time,
$W_{\text{axion} \atop \text{0-mode}}$ is parity-odd, and its flat space limit is given by Eq. (\ref{GhostNumber3Odd}). Both elements of $E_2^{2,2}$ are parity-even, and the simplest possibility is that $d_2 W_{\text{dilaton} \atop \text{0-mode}}$ equals one of them.

On the other hand, among the elements $E_2^{3,1}$, two are parity-odd and one is parity-even, as we will now explain. This means that $d_3 W_{\text{axion} \atop \text{0-mode}}$ must be equal to one of the two parity-odd elements of $E_2^{3,1}$.

\subsection[Parity of elements of \texorpdfstring{$E_2^{3,1}$}{E2(3,1)}]{Parity of elements of \texorpdfstring{$\boldsymbol{E_2^{3,1}}$}{E2(3,1)}} \label{sec:ParityOfE31}

First, let us prove that $\Lambda\circ \mu_1$ is parity-even. The explicit formula is
\begin{align*}
    \Lambda\circ\mu_1
        &= \operatorname{Str} \left(
                g^{-1}\epsilon \, (\lambda_{\text{L}}-\lambda_{\text{R}}) \, g \;
                [\Sigma, \gamma_{\text{L}}^{\alpha} \, s^{\text{L}}_{\alpha} + \gamma_{\text{R}}^{\alpha} \, s^{\text{R}}_{\alpha}]
                \right)
        \\
        &= \mathrm{i} \lambda_{\text{L}}^{\alpha} F^{-1}_{\alpha\beta} \gamma_{\text{L}}^{\beta}
        + \mathrm{i} \lambda_{\text{R}}^{\alpha} F^{-1}_{\alpha\beta} \gamma_{\text{R}}^{\beta} ,
\end{align*}
where
\begin{align*}
    \lambda_{\text{L}}^{\alpha}
        &\equiv \left(g^{-1}(\lambda_{\text{L}} - \lambda_{\text{R}}) \, g\right)_{\text{L}}^{\alpha} ,
    \\
    \lambda_{\text{R}}^{\beta}
        &\equiv \left(g^{-1}(\lambda_{\text{L}} - \lambda_{\text{R}})\, g\right)_{\text{R}}^{\beta} .
\end{align*}
Under the parity map, we have
\begin{equation*}
    \bigl( \lambda_{\text{L}}^{\alpha} , 
        \lambda_{\text{R}}^{\alpha} ,
        \gamma^{\alpha}_{\text{L}} ,
        \gamma^{\alpha}_{\text{R}} , F \bigr)
    \stackrel{R}{\longmapsto}
    \bigl( -\lambda_{\text{R}}^{\alpha} ,
        -\lambda_{\text{L}}^{\alpha} , 
        \gamma^{\alpha}_{\text{R}} , 
        \gamma^{\alpha}_{\text{L}} , -F \bigr) .
\end{equation*}
Overall, $\Lambda\circ\mu_1$ is parity-even. Since, $c_2$ is parity-odd (see Eq. (\ref{CentralChargeInFlatSpaceNotations})), the product $c_2(\Lambda\circ\mu_1)$ is also parity-odd.

To figure out the parity of $\Lambda\circ\mu^\uparrow_3$ and $\Lambda\circ\mu^\downarrow_3$, we use the notations of
Section \ref{MatrixRealization} and write
\begin{equation*}
    \Lambda\circ\mu_3^\uparrow
    = \bigl|\bigl|
    (\gamma_{\text{L}} - \mathrm{i}\gamma_{\text{R}})
    \cap (\gamma_{\text{L}} - \mathrm{i}\gamma_{\text{R}})
    \cup (\gamma_{\text{L}} - \mathrm{i}\gamma_{\text{R}})
    \cap (\lambda_{\text{L}} - \mathrm{i}\lambda_{\text{R}})
    \bigr|\bigr| ;
\end{equation*}
then, Eq. (\ref{ParityOnMatrix}) implies that the parity transformation exchanges
\begin{equation*}
    \Lambda\circ\mu^\uparrow_3 \longleftrightarrow \Lambda\circ\mu^\downarrow_3 .
\end{equation*}

\subsection[Matching \texorpdfstring{$E^{p+q=4}_{\infty}$ and $\widetilde{E}^{p+q=4}_{\infty}$}{E∞(p+q=4) and Ẽ∞(p+q=4)}]{Matching \texorpdfstring{$\boldsymbol{E^{p+q=4}_{\infty}}$ and $\boldsymbol{\widetilde{E}^{p+q=4}_{\infty}}$}{E∞(p+q=4) and Ẽ∞(p+q=4)}} \label{sec:MatchingEinfty}

We conclude that both $\bigoplus_{p=0}^4E^{p,4-p}_{\infty}$ and $\bigoplus_{p=0}^4\widetilde{E}^{p,4-p}_{\infty}$ are six-dimensional.

\section{Constraints from matching \texorpdfstring{$\boldsymbol{E}$ with $\boldsymbol{\widetilde{E}}$}{E and Ẽ}} \label{ConstraintsFromMatching}

Our description of $E_2$ in Sections \ref{Grids} and \ref{KnightMove} was only at the level of conjecture, since we have not rigorously proven the following:
\begin{equation}
    \dim E_2^{0,3} \stackrel{?}{=} 2 ,
    \quad
    \dim E_2^{1,2} \stackrel{?}{=} 0 ,
    \quad
    \dim E_2^{1,3} \stackrel{?}{=} 0 ,
    \quad
    \dim E_2^{2,2} \stackrel{?}{=} 2 .
\end{equation}
But, by matching the two spectral sequences $E_{\infty}$ with $\widetilde{E}_{\infty}$, we get some constraints on these cohomology groups, which we will now describe.

Notice that all classes we have listed in $E_2^{p+q=4}$ are manifestly covariant. Therefore all $d_{\geq 2}$ vanish on $E_2^{p+q=4}$.

Since $W_{\text{dilaton} \atop \text{0-mode}}$ and $W_{\text{axion} \atop \text{0-mode}}$ cannot be chosen covariant, we know that both of them should be killed by some $d_{\geq 2}$, and thus $E_{\infty}^{p+q=3}$ is one-dimensional and generated by $c_3$.

From matching with $\dim \widetilde{E}_2^{p+q=3} = 1$ and $\dim\widetilde{E}_2^{p+q=4} = 6$, we should have
\begin{equation} \label{ConstraintOnE2}
    \dim E_2^{0,3} + \dim E_2^{1,2}
        = \dim E_2^{2,2} + \dim E_2^{1,3} .
\end{equation}
Assuming Eq. (\ref{Gh3VsSUGRA}),
\begin{equation*}
    \dim E_2^{1,3}
        = \dim H^1(\mathfrak{g},H^3_Q)
        = \dim H^1(\mathfrak{g},\text{SUGRA}) ,
\end{equation*}
and Eq. (\ref{Gh2VsSUGRA}),
\begin{equation*}
    \dim H^1 (\mathfrak{g},\text{SUGRA})
        = \dim H^1 (\mathfrak{g},H^2_Q)
        = \dim E_2^{1,2} ,
\end{equation*}
it follows from Eq. (\ref{ConstraintOnE2}) that
\begin{equation*}
    \dim E_2^{0,3}
        = \dim E_2^{2,2} .
\end{equation*}

\section{Case \texorpdfstring{$\boldsymbol{\mathcal{R} = \mathrm{ad}}$}{cal(R) = ad}} \label{RAd}

The next simplest case is $\mathcal{R} = \mathrm{ad}$. As before, we first present the grids for the spectral sequences, and we describe the computation afterwards.

\begin{table}[h]
\centering
\renewcommand{\arraystretch}{1.5}
\begin{tabular}{|c||*{5}{>{\centering\arraybackslash}m{2.5cm}|}}
    \hline
    $\boldsymbol{4}$
        & $0$ & $0$ & $0$ & $0$ & $0$
    \\ \hline
    $\boldsymbol{3}$
        & $\operatorname{Hom}_\mathfrak{g}(\mathrm{ad},H_Q^3)$
         & $\operatorname{Ext}_\mathfrak{g}^1(\mathrm{ad},H_Q^3)$
          & $\operatorname{Ext}_\mathfrak{g}^2(\mathrm{ad},H_Q^3)$
           & $\operatorname{Ext}_\mathfrak{g}^3(\mathrm{ad},H_Q^3)$
            & $\operatorname{Ext}_\mathfrak{g}^4(\mathrm{ad},H_Q^3)$
    \\ \hline
    $\boldsymbol{2}$
        & $\operatorname{Hom}_\mathfrak{g}(\mathrm{ad},H_Q^2)$
         & $\operatorname{Ext}_\mathfrak{g}^1(\mathrm{ad},H_Q^2)$
          & $\operatorname{Ext}_\mathfrak{g}^2(\mathrm{ad},H_Q^2)$
           & $\operatorname{Ext}_\mathfrak{g}^3(\mathrm{ad},H_Q^2)$
            & $\operatorname{Ext}_\mathfrak{g}^4(\mathrm{ad},H_Q^2)$
    \\ \hline
    $\boldsymbol{1}$
        & $\mathbb{C}$
         & $0$
          & $\operatorname{Ext}_\mathfrak{g}^2(\mathrm{ad},\mathrm{ad})$
           & $\operatorname{Ext}_\mathfrak{g}^3(\mathrm{ad},\mathrm{ad})$
            & $\operatorname{Ext}_\mathfrak{g}^4(\mathrm{ad},\mathrm{ad})$
    \\ \hline
    $\boldsymbol{0}$
        & $0$
         & $\mathbb{C}$
          & $0$
           & $H^3(\mathfrak{g},\mathrm{ad})$
            & $\operatorname{Ext}_\mathfrak{g}^4(\mathrm{ad},\mathbb{C})$
    \\ \hline\hline
    \diagbox[
        dir=SW,
        innerwidth=0.75cm, 
        innerleftsep = 0.125cm,
        innerrightsep = 0.125cm,
        height=1.75\line,
        ]{$\boldsymbol{q}$}{$\boldsymbol{p}$}
        & $\boldsymbol{0}$
         & $\boldsymbol{1}$ 
          & $\boldsymbol{2}$ 
           & $\boldsymbol{3}$ 
            & $\boldsymbol{4}$
    \\ \hline
\end{tabular}
\caption{The spaces in the corner of the second page $E_2^{p,q} = \operatorname{Ext}_{\mathfrak{g}}^p \bigl( \mathrm{ad} , H^q_Q(\operatorname{Coind}^G_H \mathcal{P}^\bullet) \bigr)$.}
\label{tab:E2ad}
\end{table}

\begin{table}[!ht]
\centering
\renewcommand{\arraystretch}{1.5}
\begin{tabular}{|c||*{5}{>{\centering\arraybackslash}m{2.5cm}|}}
    \hline
    $\boldsymbol{4}$
        & $0$ & $0$ & $0$ & $0$ & $0$
    \\ \hline
    $\boldsymbol{3}$
        & $0$
         & \scalebox{.65}{$H^3(\mathfrak{h}) \otimes {}$} 
                \scalebox{.65}{$H_Q^1\bigl(\operatorname{Hom}_\mathfrak{h}(\mathrm{ad},\mathcal{P}^\bullet)\bigr)$}
          & $0$
           & \scalebox{.65}{$H^3(\mathfrak{h}) \otimes {} $} 
                \scalebox{.65}{$H_Q^3\bigl(\operatorname{Hom}_\mathfrak{h}(\mathrm{ad},\mathcal{P}^\bullet)\bigr)$}
            & $0$
    \\ \hline
    $\boldsymbol{2}$
        & $0$
         & $0$
          & $0$
           & $0$
            & $0$
    \\ \hline
    $\boldsymbol{1}$
        & $0$
         & $0$
          & $0$
           & $0$
            & $0$
    \\ \hline
    $\boldsymbol{0}$
        & $0$
         & \scalebox{.65}{$H_Q^1\bigl( \operatorname{Hom}_\mathfrak{h}(\mathrm{ad},\mathcal{P}^\bullet) \bigr)$}
          & $0$
           & \scalebox{.65}{$H_Q^3\bigl( \operatorname{Hom}_\mathfrak{h}(\mathrm{ad},\mathcal{P}^\bullet) \bigr)$}
            & \scalebox{.65}{$H_Q^4\bigl( \operatorname{Hom}_\mathfrak{h}(\mathrm{ad},\mathcal{P}^\bullet) \bigr)$}
    \\ \hline\hline
    \diagbox[
        dir=SW,
        innerwidth=0.75cm, 
        innerleftsep = 0.125cm,
        innerrightsep = 0.125cm,
        height=1.75\line,
        ]{$\boldsymbol{q}$}{$\boldsymbol{p}$}
        & $\boldsymbol{0}$
         & $\boldsymbol{1}$ 
          & $\boldsymbol{2}$ 
           & $\boldsymbol{3}$ 
            & $\boldsymbol{4}$
    \\ \hline
\end{tabular}
\caption{The spaces in the corner of the second page $\widetilde{E}_2^{p,q} = H_Q^p\bigl( \operatorname{Ext}_{\mathfrak{g}}^q ( \mathrm{ad} , \operatorname{Coind}^\mathfrak{g}_\mathfrak{h} \mathcal{P}^\bullet) \bigr)$.}
\label{tab:E2tildead}
\end{table}
\begin{table}[!ht]
\centering
\renewcommand{\arraystretch}{1.5}
\begin{tabular}{|c||*{5}{>{\centering\arraybackslash}m{2.5cm}|}}
    \hline
    $\boldsymbol{4}$
        & $0$ & $0$ & $0$ & $0$ & $0$
    \\ \hline
    $\boldsymbol{3}$
        & $0$
         & ${\color{gray} \operatorname{Ext}_\mathfrak{g}^1(\mathrm{ad},H_Q^3)}$
          & ${\color{gray} \operatorname{Ext}_\mathfrak{g}^2(\mathrm{ad},H_Q^3)}$
           & ${\color{gray} \operatorname{Ext}_\mathfrak{g}^3(\mathrm{ad},H_Q^3)}$
            & ${\color{gray} \operatorname{Ext}_\mathfrak{g}^4(\mathrm{ad},H_Q^3)}$
    \\ \hline
    $\boldsymbol{2}$
        & $0$
         & $\bullet\, \Lambda\bigl\langle [\Sigma,c] \bigr\rangle \, \Lambda$
          & ${\color{gray} \operatorname{Ext}_\mathfrak{g}^2(\mathrm{ad},H_Q^2)}$
           & ${\color{gray} \operatorname{Ext}_\mathfrak{g}^3(\mathrm{ad},H_Q^2)}$
            & ${\color{gray} \operatorname{Ext}_\mathfrak{g}^4(\mathrm{ad},H_Q^2)}$
    \\ \hline
    $\boldsymbol{1}$
        & $\mathbb{C}$
         & $0$
          & $\bullet\, \scalebox{.75}{$c_2 \,\mathrm{id}$}$, $\bullet\, \scalebox{.55}{ $\Lambda\langle[\Sigma,c]\rangle \operatorname{STr}( [\Sigma,c]\,x)$}$
           & ${\color{gray} \operatorname{Ext}_\mathfrak{g}^3(\mathrm{ad},\mathrm{ad})}$
            & ${\color{gray} \operatorname{Ext}_\mathfrak{g}^4(\mathrm{ad},\mathrm{ad})}$
    \\ \hline
    $\boldsymbol{0}$
        & $0$
         & $\mathbb{C}$
          & $0$
           & $\mathbb{C}^3$
            & ${\color{gray} \operatorname{Ext}_\mathfrak{g}^4(\mathrm{ad},\mathbb{C})}$
    \\ \hline\hline
    \diagbox[
        dir=SW,
        innerwidth=0.75cm, 
        innerleftsep = 0.125cm,
        innerrightsep = 0.125cm,
        height=1.75\line,
        ]{$\boldsymbol{q}$}{$\boldsymbol{p}$}
        & $\boldsymbol{0}$
         & $\boldsymbol{1}$ 
          & $\boldsymbol{2}$ 
           & $\boldsymbol{3}$ 
            & $\boldsymbol{4}$
    \\ \hline
\end{tabular}
    \caption{Generators in the corner of the second page $E_2^{p,q} = \operatorname{Ext}_{\mathfrak{g}}^p \bigl( \mathrm{ad} , H^q_Q(\operatorname{Coind}^G_H \mathcal{P}^\bullet) \bigr)$.}
\label{tab:E2adgen}
\end{table}

\begin{table}[h]
\centering
\renewcommand{\arraystretch}{1.5}
\begin{tabular}{|c||*{5}{>{\centering\arraybackslash}m{2.5cm}|}}
    \hline
    $\boldsymbol{4}$
        & $0$ & $0$ & $0$ & $0$ & $0$
    \\ \hline
    $\boldsymbol{3}$
        & $0$
         & \scalebox{.65}{$\color{gray} H^3(\mathfrak{h}) \otimes {}$}
            \scalebox{.65}{$\color{gray} H_Q^1\bigl(\operatorname{Hom}_\mathfrak{h}(\mathrm{ad},\mathcal{P}^\bullet)\bigr)$}
          & $0$
           & \scalebox{.65}{$\color{gray} H^3(\mathfrak{h}) \otimes {} $}
            \scalebox{0.65}{$\color{gray} H_Q^3\bigl(\operatorname{Hom}_\mathfrak{h}(\mathrm{ad},\mathcal{P}^\bullet)\bigr)$}
            & $0$
    \\ \hline
    $\boldsymbol{2}$
        & $0$
         & $0$
          & $0$
           & $0$
            & $0$
    \\ \hline
    $\boldsymbol{1}$
        & $0$
         & $0$
          & $0$
           & $0$
            & $0$
    \\ \hline
    $\boldsymbol{0}$
        & $0$
         & $\bullet\, \scalebox{.75}{$\Lambda$}$,
            $\bullet\, \scalebox{.5}{$\operatorname{STr}\bigl( xg^{-1} \bigl[\Sigma,(\lambda_{\text{L}}+\lambda_{\text{R}})\bigr]\bigr)$}$
          & $0$
           & \scalebox{.85}{$\mathbb{C}^6$,} \scalebox{.85}{see Eq. \eqref{E2tildead30generators}}
            & \scalebox{0.65}{$\color{gray} H_Q^4\bigl( \operatorname{Hom}_\mathfrak{h}(\mathrm{ad},\mathcal{P}^\bullet) \bigr)$}
    \\ \hline\hline
    \diagbox[
        dir=SW,
        innerwidth=0.75cm, 
        innerleftsep = 0.125cm,
        innerrightsep = 0.125cm,
        height=1.75\line,
        ]{$\boldsymbol{q}$}{$\boldsymbol{p}$}
        & $\boldsymbol{0}$
         & $\boldsymbol{1}$ 
          & $\boldsymbol{2}$ 
           & $\boldsymbol{3}$ 
            & $\boldsymbol{4}$
    \\ \hline
\end{tabular}
    \caption{Generators in the corner of the second page $\widetilde{E}_2^{p,q} = H^q_{\mathfrak{h}}(\mathbb{C})\otimes H^p_Q\left(\mbox{Hom}_{\mathfrak{h}}\left(\mbox{ad},{\cal P}^{\bullet}\right)\right)$.}
\label{tab:E2tildeadgen}
\end{table}

\subsection[Computation of \texorpdfstring{$E_2^{p,3-p}$}{E2(p,3-p)}]{Computation of \texorpdfstring{$\boldsymbol{E_2^{p,3-p}}$}{E2(p,3-p)}} 

The generators of $E_2^{p,3-p}$ are listed in the grid table \ref{tab:E2adgen}. Notice that all the generators are covariant, thus the higher differentials are zero. The dimensions of those spaces are
\begin{equation*}
  \dim E_2^{1,2} = 1 , \quad
  \dim E_2^{2,1} = 2 , \quad
  \dim E_2^{3,0} = 3 ;
\end{equation*}
the total dimension is six, matching the dimension of the corresponding sector of $\widetilde{E}_2$.

\subsection[\texorpdfstring{$\widetilde{E}_2^{p,0}$}{Ẽ2(p,0)} and the covariant pure spinor complex]{\texorpdfstring{$\boldsymbol{\widetilde{E}_2^{p,0}}$}{Ẽ2(p,0)} and the covariant pure spinor complex} \label{sec:EquivariantPureSpinorComplex}

Let us restrict the pure spinor complex on the subspace of cochains which transform in the adjoint representation of $\mathfrak{g}$. Its cohomology is identified with $\widetilde{E}_2^{p,0}$ --- the bottom rows of tables \ref{tab:E2tildead} and \ref{tab:E2tildeadgen}.

It is immediate that the zeroth cohomology is trivial. The first cohomology is two-dimensional, and it is generated by the elements
\begin{align*}
    \operatorname{Str}\bigl(x g^{-1}(\lambda_{\text{L}} - \lambda_{\text{R}}) \, g \bigr) = \Lambda ,
    \\
    \operatorname{Str}\bigl(x g^{-1}[\Sigma, (\lambda_{\text{L}} + \lambda_{\text{R}})] \, g \bigr) .
\end{align*}
The second cohomology, however, is trivial.

The third cohomology is six-dimensional, generated by the elements
\begin{subequations} \label{E2tildead30generators}
\begin{gather}
    \operatorname{Str}(\lambda_{\text{L}}\lambda_{\text{R}}) \, \Lambda ,
    \\
    \operatorname{Str}( \lambda_{\text{L}}\lambda_{\text{R}} ) \, \operatorname{Str}\bigl(xg^{-1}[\Sigma, (\lambda_{\text{L}} + \lambda_{\text{R}})] \, g \bigr) ,
\end{gather}
and the two pairs
\begin{gather}
    \lambda_{\text{L/R}}^2 \, \Lambda ,
    \\
    \lambda_{\text{L/R}}^2 \, \operatorname{Str}\bigl( xg^{-1}[\Sigma, (\lambda_{\text{L}} + \lambda_{\text{R}})] \, g \bigr) .
\end{gather}
\end{subequations}

\section*{Acknowledgments}\label{Acknowledments}

We want to thank Nathan Berkovits and Thiago Fleury for discussions.
This work was supported in part by FAPESP grant {2019/21281-4}, and in part by FAPESP grant {2021/14335-0}, and in part by CNPq grant ``Produtividade em Pesquisa'' {307191/2022-2}.

TOF is grateful for the hospitality of Perimeter Institute where part of this work was carried out. Research at Perimeter Institute is supported in part by the Government of Canada through the Department of Innovation, Science and Economic Development and by the Province of Ontario through the Ministry of Colleges and Universities.

\bibliographystyle{jhep} \renewcommand{\refname}{Bibliography}
\addcontentsline{toc}{section}{Bibliography}
\bibliography{andrei.bib}

\end{document}